\documentclass[pra,aps,twocolumn,superscriptaddress,floatfix,noshowpacs,nofootinbib,10pt]{revtex4-2}

\usepackage[utf8]{inputenc}
\usepackage{microtype}
\usepackage{amsmath,amssymb}
\usepackage{graphicx}
\usepackage[colorlinks=true]{hyperref}

\usepackage{bm}
\renewcommand{\vec}{\bm}
\renewcommand{\Re}{\operatorname{Re}}
\renewcommand{\Im}{\operatorname{Im}}

\newcommand{\dif}{\mathrm{d}}
\newcommand{\mi}{\mathrm{i}}
\newcommand{\me}{\mathrm{e}}

\begin{document}

\title{Spin-orbit-coupled fluids of light in bulk nonlinear media}

\author{Giovanni I. Martone}
\email{giovanni\textunderscore italo.martone@lkb.upmc.fr}
\affiliation{Laboratoire Kastler Brossel,
Sorbonne Universit\'{e}, CNRS, ENS-PSL Research University, 
Coll\`{e}ge de France; 4 Place Jussieu, 75005 Paris, France}

\author{Tom Bienaim\'{e}}
\email{tom.bienaime@lkb.ens.fr}
\affiliation{Laboratoire Kastler Brossel,
Sorbonne Universit\'{e}, CNRS, ENS-PSL Research University, 
Coll\`{e}ge de France; 4 Place Jussieu, 75005 Paris, France}

\author{Nicolas Cherroret}
\email{cherroret@lkb.upmc.fr}
\affiliation{Laboratoire Kastler Brossel,
Sorbonne Universit\'{e}, CNRS, ENS-PSL Research University, 
Coll\`{e}ge de France; 4 Place Jussieu, 75005 Paris, France}

\begin{abstract}
We show that nonparaxial polarized light beams propagating in a bulk nonlinear Kerr medium naturally exhibit a coupling between
the motional and the polarization degrees of freedom, realizing a spin-orbit-coupled mixture of fluids of light. We investigate the impact of this
mechanism on the Bogoliubov modes of the fluid, using a suitable density-phase formalism built upon a linearization of the exact
Helmholtz equation. The Bogoliubov spectrum is found to be anisotropic, and features both low-frequency gapless branches and
high-frequency gapped ones. We compute the amplitudes of these modes and propose a couple of experimental protocols to study
their excitation mechanisms. This allows us to highlight a phenomenon of hybridization between density and spin modes, which is absent
in the paraxial description and represents a typical fingerprint of spin-orbit coupling.
\end{abstract}

\maketitle

\section{Introduction}
\label{sec:introduction}
Quantum fluids of light represent a novel and flexible kind of many-body system, whose constituents are photons that can effectively interact
with each other (see~\cite{Carusotto2013} for a review). They make it possible to observe in optics many phenomena usually encountered
in cold-atom systems, such as condensation~\cite{Kasprzak2006,Klaers2010}, superfluidity~\cite{Amo2009,Lerario2017}, or nucleation of nonlinear
excitations~\cite{Amo2011,Nardin2011,Sanvitto2011}. To achieve a fluid of light, a simple approach consists in propagating a quasi-monochromatic
beam of light in a medium with sufficiently strong Kerr optical nonlinearity~\cite{Carusotto2014,Noh2016}. This setup, which does not involve any
cavity, can be modeled as a $(2+1)$-dimensional interacting system, with the two spatial dimensions lying in the transverse plane while the
propagation direction plays the role of an effective time. The third-order Kerr nonlinearity is interpreted as a photon-photon interaction. This
configuration has been experimentally realized using a variety of materials, including photorefractive crystals~\cite{Wan2007,Michel2018},
thermo-optic media~\cite{Elazar2013,Vocke2015,Vocke2016}, and hot atomic vapors~\cite{Santic2018,Fontaine2018}. In the paraxial approximation,
where the propagation occurs at a small angle with the optical axis of the medium, the effective-time evolution of the beam is described by a
nonlinear Schr\"{o}dinger equation. The latter is mathematically analogous to the Gross-Pitaevskii equation governing the real-time dynamics of an
atomic Bose-Einstein condensate. Consequently, small oscillating fluctuations of the beam intensity field on top of a fixed background are described
by the standard Bogoliubov theory~\cite{Bogoliubov1947,Castin_review,Pethick_Smith_book,Pitaevskii_Stringari_book}, as was recently
confirmed experimentally~\cite{Vocke2015,Fontaine2018}. Large perturbations on top of a small background on the other hand lead to the creation of
dispersive shock waves~\cite{Bienaime2021,Abuzarli2021,Isoard2019,Ivanov2020}. Recently, this platform was also exploited to explore the generation
of topological defects and the associated turbulence~\cite{Eloy2020,Rodrigues2020}, analogue cosmological Sakharov oscillations in the
density-density correlations of a quantum fluid of light~\cite{Steinhauer2021,Larre2015}, as well as the spontaneous emergence of long-range
coherence~\cite{Fusaro2017,BardonBrun2020}.

In recent times, multicomponent atomic quantum gases with spin-orbit coupling have become a very active field of research (see the reviews
in~\cite{Dalibard2011review,Galitski2013review,Zhou2013review,Goldman2014review,Zhai2015review,Li2015review,Zhang2016review} and
references therein). These systems are engineered in the laboratory by coupling the atoms with properly designed laser fields. On the one
hand, they enable one to study phenomena and systems typical of solid-state physics (spin-Hall effect, Majorana fermions, topological insulators)
in highly controllable setups. On the other hand, they allow for the realization of novel many-body configurations, with no counterpart in other
domains of condensed matter physics. In this context, a natural question is whether the physics of spin-orbit coupling can also be investigated in the
framework of fluids of light. It is known that -- in contrast to atomic gases -- light propagating in inhomogeneous media is naturally subject to an intrinsic
spin-orbit interaction, which is predicted by Maxwell theory~\cite{Bliokh2015,Ling2017}. Such a mechanism has been observed in a number of optical
configurations involving light transmitted or reflected at dielectric interfaces~\cite{Hosten08}, plasmonic slits~\cite{Gorodetski2012}, nonparaxial
beams~\cite{Zhao2009}, or light propagating in random media~\cite{Cherroret2018, BardonBrun2019}. In the context of fluids of light, signatures
of spin-orbit interactions and of spin Hall effects were reported in microcavity exciton-polaritons~\cite{Leyder2007,Tercas2014,Nalitov2015,Sala2015}.
In the present work, we show that a spin-orbit interaction naturally occurs for fluids of light in the cavityless propagating geometry discussed above.
For that purpose, we construct a general formalism for spin mixtures of fluids of light that goes beyond the paraxial approximation in which, by
construction, any coupling between polarization and orbital motion is discarded. By linearizing the full nonlinear Helmholtz equation describing
the system and computing the Bogoliubov modes, we show that the excitation spectrum exhibits features, like the anisotropy of the frequencies and
the hybridization of the density and spin modes, that are typical of spin-orbit-coupled atomic Bose gases~\cite{Martone2012,Liao2013,Geier2021}.
We also discuss specific experimental setups aiming to detect these phenomena.

This article is organized as follows. Section~\ref{sec:model} is devoted to the presentation of the physical model describing our nonparaxial fluid
of light, based on the nonlinear vector Helmholtz equation, with emphasis on the phenomenon of spin-orbit coupling of light.
In Sec.~\ref{sec:dens_phase} we reformulate this model in terms of appropriately chosen density and phase variables and derive the associated Lagrangian.
The core elements of our theory are presented in Sec.~\ref{sec:bogo_theory}: here we linearize the field equations about a uniform background configuration
and study the properties of their solutions, i.e., the Bogoliubov modes. The particular case of the paraxial limit is discussed in Sec.~\ref{sec:paraxial_lim},
where we show that the system behaves like a standard binary mixture of Bose-Einstein condensates, with a density and a spin branch in its
excitation spectrum. Sections~\ref{sec:bogo_spectrum} and~\ref{sec:hybr_mode} contain the main predictions of our formalism:
in Sec.~\ref{sec:bogo_spectrum} we derive the exact Bogoliubov spectrum of the system, pointing out the presence of multiple branches and
their anisotropy. In Sec.~\ref{sec:hybr_mode} we demonstrate the existence of a phenomenon of mode hybridization due to spin-orbit-coupling
of light in a fluid mixture, and we propose two concrete experimental scenarios where it can be observed. We finally summarize our findings
in Sec.~\ref{sec:conclusion}. Further technical details are collected in two Appendices.

\section{The model}
\label{sec:model}
We start our analysis by describing the physical setup under consideration in Sec.~\ref{subsec:model_setup}. In Sec.~\ref{subsec:model_lag_soc}
we introduce the Lagrangian of the model and discuss the connection with the phenomenon of spin-orbit coupling.

\subsection{Setup description}
\label{subsec:model_setup}
Let us consider a monochromatic light beam propagating in a dielectric material with cubic Kerr nonlinearity. We write the electric field
as $\vec{E}(\vec{r},t) = \Re \left[ \vec{\mathcal{E}}(\vec{r}) \me^{- \mi \omega_0 t} \right]$, where $\omega_0$ is the field frequency and
$\vec{\mathcal{E}}$ its complex amplitude. The components of $\vec{\mathcal{E}}$ inside the medium obey the so-called Helmholtz equation
\begin{equation}
- \nabla^2 \vec{\mathcal{E}} + \nabla \left( \nabla \cdot \vec{\mathcal{E}} \right) - \beta_0^2 \vec{\mathcal{E}}
+ 2 \beta_0 g_I |\vec{\mathcal{E}}|^2 \vec{\mathcal{E}} + 2 \beta_0 g_P \vec{\mathcal{E}}^2 \vec{\mathcal{E}}^*
= 0 \, .
\label{eq:helm_eq}
\end{equation}
Here $\beta_0 = n_0 \omega_0 / c$ is the propagation constant, with $n_0$ the linear refractive index and $c$ the vacuum speed of light. Notice
that two distinct nonlinear terms are present: one is proportional to the total optical intensity $|\vec{\mathcal{E}}|^2$, the other to the square
field $\vec{\mathcal{E}}^2$ (which in turn depends on the field polarization). The corresponding coupling strengths are $g_{I(P)} = - n_{2,I(P)}
\omega_0 / c$, where $n_{2,I(P)}$ denotes the two nonlinear refractive indices. The appearance of these two kinds of nonlinearity is a consequence
of the formal properties of the cubic susceptibility of isotropic materials (see, e.g., Refs.~\cite{Landau_Lifshitz_08_book,Agrawal_book}). A simple
physical interpretation can be formulated thinking in analogy with spin-$1$ bosonic systems undergoing rotationally-invariant $s$-wave collisions:
the strength of the interaction can be different for colliding pairs of total spin $0$ and $2$, thus justifying the need for two independent coupling
constants~\cite{StamperKurn2013review}. If both $g_d$ and $g_s$ are nonzero, the two nonlinear terms in Eq.~\eqref{eq:helm_eq} only
reduce to a single one in two cases: when the electric field is linearly polarized and one has $|\vec{\mathcal{E}}|^2 \vec{\mathcal{E}} =
\vec{\mathcal{E}}^2 \vec{\mathcal{E}}^*$, or when it is circularly polarized and thus $\vec{\mathcal{E}}^2 = 0$. In the present work we consider fields
of arbitrary polarization, hence one has to include both nonlinear couplings in the Helmholtz equation~\eqref{eq:helm_eq}.

The main goal of our analysis is to investigate the effects of the spin-orbit coupling of light, which is encoded in the second term on the left-hand side
of Eq.~\eqref{eq:helm_eq}. This term is absent in linear homogeneous media, where, according to Maxwell's equations, $\nabla \cdot \vec{\mathcal{E}}$
vanishes. Taking the divergence on both sides of Eq.~\eqref{eq:helm_eq} one can see that the spin-orbit coupling in our system originates from the spatial
variations of the nonlinear terms [see also Eq.~\eqref{eq:paraxial_field_div} below]. This motivates the Bogoliubov-like approach we will develop in
this paper: by imprinting a small fluctuation on top of a uniform background, one will trigger a spin-orbit coupling accessible through perturbation theory.

For concreteness we consider a material that extends infinitely along the $\vec{r}_\perp = (x,y)$ plane, as well as along the positive direction
of the optical axis $z$. We assume that at the $z=0$ interface between the air and the medium, the latter is illuminated by a laser of given
electric-field profile $\vec{\mathcal{E}}(\vec{r}_\perp,z\!=\!0)$ (see Fig.~\ref{fig:hybr_configs} below and the related discussion of
Sec.~\ref{sec:hybr_mode} for practical examples of this setting). The components of the field inside the medium are then found by solving the three
coupled differential equations~\eqref{eq:helm_eq} with the above condition. This is in fact an initial-value problem, in which the $z$ coordinate
can be regarded as an effective time. Since Eq.~\eqref{eq:helm_eq} is of second order in $z$, one also has to specify the first derivative of the field
profile at the interface, $\dot{\vec{\mathcal{E}}}(\vec{r}_\perp,z\!=\!0)$ (we use the standard dot notation for writing derivatives with respect to $z$).

An important property of Eq.~\eqref{eq:helm_eq} is that its projection onto the $z$ axis depends on the longitudinal field component $\mathcal{E}_z$
but not on the effective-time derivatives $\dot{\mathcal{E}}_z$ and $\ddot{\mathcal{E}}_z$. This means that one can eliminate $\mathcal{E}_z$ by
expressing it as a function of the transverse components of $\vec{\mathcal{E}}$ and their derivatives (in other words, $\mathcal{E}_z$
is not an independent dynamical variable of our problem). However, in practice this elimination can be difficult to perform in the presence of nonlinearities,
unless one considers the linearized version of Eq.~\eqref{eq:helm_eq}, as we will do in Sec.~\ref{sec:bogo_theory}.

\subsection{Lagrangian formulation and spin-orbit coupling}
\label{subsec:model_lag_soc}
Instead of working directly with the vector equation~\eqref{eq:helm_eq}, we develop our analysis starting from the corresponding Lagrangian, which is
a scalar quantity and is thus easier to manipulate. It reads $L = \int \dif^2 r_\perp \, \mathcal{L}$, where the Lagrangian density is
\begin{equation}
\begin{split}
\mathcal{L}
= {}&{} - \frac{1}{2 \beta_0}
\left\{ \big[ \left( \vec{S} \cdot \nabla \right)_{ij} \mathcal{E}_j \big]^* \big[ \left( \vec{S} \cdot \nabla \right)_{ij'} \mathcal{E}_{j'} \big]
- \beta_0^2 \mathcal{E}_i^* \mathcal{E}_i \right\} \\
&{} - \frac{g_d}{2} \mathcal{E}_i^* \mathcal{E}_j^* \mathcal{E}_i \mathcal{E}_j
- \frac{g_s}{2} \mathcal{E}_i^* \mathcal{E}_{i'}^* (S_k)_{ij} (S_k)_{i'j'} \mathcal{E}_j \mathcal{E}_{j'}.
\label{eq:helm_lagr}
\end{split}
\end{equation}
Here and henceforth Latin indices $i,j,k,\ldots$ take values $x,y,z$ (or $+,-,z$ if one works in the circular basis, see below), and we implicitly sum over
repeated indices unless otherwise specified. The prefactor of $\mathcal{L}$ has been chosen such that it reduces to the Gross-Pitaevskii Lagrangian
density in the paraxial limit, see Eq.~\eqref{eq:paraxial_lagr} below. Here, the components $S_k$ of the spin operator are $3 \times 3$ matrices.
Correspondingly, the spin optical intensity is defined as the vector with components $\mathcal{E}_i^* (S_k)_{ij} \mathcal{E}_j$. The two interaction terms
in $\mathcal{L}$ have strengths $g_d = g_I + g_P$ and $g_s = - g_P$. They are proportional to the square of the total optical intensity and the spin optical
intensity, respectively. This is the most general form of the two-body contact interaction for a three-component system with full rotational
invariance~\cite{StamperKurn2013review}.

Notice that the kinetic term of the Lagrangian density~\eqref{eq:helm_lagr} has the peculiar form $(\vec{S} \cdot \nabla)^2$, and thus features
a three-dimensional and fully isotropic spin-orbit coupling. This is a different situation compared to the one arising in atomic gases, where the spin-orbit coupling
is often taken as a combination of the Rashba~\cite{Bychkov1984} and Dresselhaus~\cite{Dresselhaus1955} terms, which are linear in the
particle's spin and momentum. In addition, in those systems, the spin-orbit interaction can be anisotropic and involves only a subset of spatial
directions (see the reviews in~\cite{Dalibard2011review,Galitski2013review,Zhou2013review,Goldman2014review,Zhai2015review,Li2015review,
Zhang2016review} for further information).

Equation~\eqref{eq:helm_lagr} does not depend on the basis in which $\vec{\mathcal{E}}$ is expressed. If one chooses the Cartesian basis, the entries of $S_k$
simply read $(S_k)_{ij} = - \mi \varepsilon_{ijk}$. Taking the Euler-Lagrange equation $\partial_i \partial \mathcal{L}/\partial(\partial_i \mathcal{E}_j^*)
- \partial \mathcal{L}/\partial \mathcal{E}_j^* = 0$ and using the tensor identity $\varepsilon_{ijk} \varepsilon_{i'j'k} = \delta_{ii'} \delta_{jj'}
- \delta_{ij'} \delta_{ji'}$, one eventually recovers the Helmholtz equation~\eqref{eq:helm_eq}. However, in the following we will use the basis of
\emph{circular polarizations}, which is a natural choice when dealing with spin-related phenomena. It is spanned by the three unit vectors
$\hat{\vec{e}}_\pm = \mp (\hat{\vec{e}}_x \pm \mi \hat{\vec{e}}_y) / \sqrt{2}$ and $\hat{\vec{e}}_z$ (we take $z$ as the quantization axis of the angular momentum).
The components of the complex electric field in this basis are $\mathcal{E}_\pm = \mp (\mathcal{E}_x \mp \mi \mathcal{E}_y) / \sqrt{2}$ and $\mathcal{E}_z$.
Regarding the spin operator $\vec{S}$, its $z$ component in the circular basis $\left\{ \hat{\vec{e}}_+ , \hat{\vec{e}}_z , \hat{\vec{e}}_- \right\}$ is diagonal,
$S_z = \operatorname{diag}(1,0,-1)$, and the other two are given by
\begin{equation}
S_x = \frac{1}{\sqrt{2}}
\begin{pmatrix}
0 & 1 & 0 \\
1 & 0 & 1 \\
0 & 1 & 0
\end{pmatrix}
\, , \quad
S_y = \frac{1}{\sqrt{2}}
\begin{pmatrix}
0 & -\mi & 0 \\
\mi & 0 & -\mi \\
0 & \mi & 0
\end{pmatrix}
\, .
\label{eq:spin_mats}
\end{equation}

\section{Density-phase formalism}
\label{sec:dens_phase}
The Helmholtz equation~\eqref{eq:helm_eq} describes the effective-time evolution of the electric field starting from an arbitrary initial profile.
Here, we consider situations where the total field can be represented in the form of a large background term with a small
fluctuation on top of it. The background is a stationary solution of Eq.~\eqref{eq:helm_eq}, while the effective-time evolution of the fluctuation
can be investigated by linearizing this equation around the background. This procedure is analogous to the Bogoliubov approach for studying fluctuations
on top of a Bose-Einstein condensate (see, e.g., the books~\cite{Pethick_Smith_book,Pitaevskii_Stringari_book} and references therein).
However, it should be kept in mind that fluids of light are two-dimensional systems: since $z$ plays the role of time, only $x$ and $y$ should
be regarded as spatial coordinates. This means that the phases of the components of $\vec{\mathcal{E}}$ can experience large fluctuations around their
background values. Consequently, one should not directly perform the expansion of the total field $\vec{\mathcal{E}}$ about the background, as it requires
both intensity and phase fluctuations to be small; rather, one should tackle this problem by means of the so-called density-phase
formalism~\cite{Popov1972,Popov_book,Mora2003,Petrov2004}. We point out that this is not a strict necessity for the calculations of the present paper:
since the observables we consider (mainly the excitation spectrum and the beam intensity) do not depend on the phases of $\vec{\mathcal{E}}$,
they could be safely evaluated using the field formalism. However, working in the density-phase framework becomes mandatory for accurately describing
other quantities such as field correlation functions~\cite{BardonBrun2020}. Because of its more general validity, in this article we will work
within this formalism, whose extension to a spin-orbit-coupled fluid of light is described in the present section.

The starting point is the Lagrangian density~\eqref{eq:helm_lagr}, which for simplicity is decomposed as $\mathcal{L} = \mathcal{L}_\perp + \mathcal{L}_z$.
$\mathcal{L}_z$ collects all the terms in $\mathcal{L}$ depending on $\mathcal{E}_z$, whereas the remaining ones are included in $\mathcal{L}_\perp$.
One has:
\begin{widetext}
\begin{equation}
\mathcal{L}_\perp = - \frac{1}{2 \beta_0} \left[ (S_z^2)_{\beta\beta'} \dot{\mathcal{E}}_\beta^* \dot{\mathcal{E}}_{\beta'} 
+ (S_\alpha S_{\alpha'})_{\beta\beta'} \partial_\alpha\mathcal{E}_\beta^* \, \partial_{\alpha'}\mathcal{E}_{\beta'} \right]
+ \frac{\beta_0}{2} \mathcal{E}_\beta^* \mathcal{E}_\beta 
- \frac{g_d}{2} \mathcal{E}_\beta^* \mathcal{E}_{\beta'}^* \mathcal{E}_\beta \mathcal{E}_{\beta'}
- \frac{g_s}{2} (S_z)_{\beta\gamma} (S_z)_{\beta'\gamma'} \mathcal{E}_{\beta}^* \mathcal{E}_{\beta'}^* \mathcal{E}_{\gamma} \mathcal{E}_{\gamma'}
\label{eq:helm_lagr_perp}
\end{equation}
and
\begin{equation}
\begin{split}
\mathcal{L}_z = {}&{} - \frac{1}{2 \beta_0} \left[ (S_\alpha S_{\alpha'})_{zz} \partial_\alpha\mathcal{E}_z^* \, \partial_{\alpha'}\mathcal{E}_z
+ (S_\alpha S_z)_{z\beta} \partial_\alpha\mathcal{E}_z^* \, \dot{\mathcal{E}}_\beta
+ (S_z S_\alpha)_{\beta z} \dot{\mathcal{E}}_\beta^* \partial_\alpha\mathcal{E}_z \right] + \frac{\beta_0}{2} |\mathcal{E}_z|^2 
- g_d \mathcal{E}_\beta^* \mathcal{E}_\beta |\mathcal{E}_z|^2 - \frac{g_d}{2} |\mathcal{E}_z|^4 \\
{}&{} - g_s (S_\alpha)_{\beta z} (S_\alpha)_{z \beta'} \mathcal{E}_\beta^* \mathcal{E}_{\beta'} |\mathcal{E}_z|^2
- \frac{g_s}{2} \big[ (S_\alpha)_{z \beta} (S_\alpha)_{z \beta'} \mathcal{E}_z^{*2} \mathcal{E}_\beta \mathcal{E}_{\beta'}
+ (S_\alpha)_{\beta z} (S_\alpha)_{\beta' z} \mathcal{E}_\beta^* \mathcal{E}_{\beta'}^* \mathcal{E}_z^2 \big] \, .
\end{split}
\label{eq:helm_lagr_z}
\end{equation}
\end{widetext}
In these expressions Greek indices take values $x,y$ or $\pm$, depending on whether one works in the Cartesian or circular basis. Notice that $\mathcal{L}_z$
does not depend on $\dot{\mathcal{E}}_z$, which is consistent with the fact that $\mathcal{E}_z$ is not an independent dynamical variable, as discussed in
Sec.~\ref{sec:model}. We now introduce the density and phase variables by parametrizing the electric field as follows:
\begin{equation}
\begin{pmatrix}
\mathcal{E}_+ \\
\mathcal{E}_z \\
\mathcal{E}_-
\end{pmatrix}
= \me^{\mi \Theta}
\begin{pmatrix}
\sqrt{I} \cos \frac{\vartheta}{2} \, \me^{\mi \chi/2} \\
\hat{\mathcal{E}}_z \\
\sqrt{I} \sin \frac{\vartheta}{2} \, \me^{- \mi \chi/2}
\end{pmatrix} \, .
\label{eq:dp_var}
\end{equation}
Here $I = |\mathcal{E}_+|^2 + |\mathcal{E}_-|^2$ is the optical intensity due to the transverse polarization components, $\vartheta \in \left[0,\pi\right]$
quantifies their relative weight and $\chi \in \left[0,2\pi\right[$ their relative phase. The complex field $\hat{\mathcal{E}}_z$ differs from the full
longitudinal component $\mathcal{E}_z$ by a phase factor, and the quantity $\Theta \in \left[0,2\pi\right[$ entering this factor can be regarded as the global
phase of the electric field. All these variables are functions of the transverse and longitudinal coordinates, $\vec{r}_\perp$ and $z$.

We now rewrite the two Lagrangian densities~\eqref{eq:helm_lagr_perp} and~\eqref{eq:helm_lagr_z} in terms of the new density-phase variables.
In doing so it is useful to recall that the entries of the spin-$1$ matrices in the circular basis [see Eq.~\eqref{eq:spin_mats}] satisfy the identities
$(S_z^2)_{\beta\beta'} = \delta_{\beta\beta'}$, $(S_\alpha S_{\alpha'})_{zz} = \delta_{\alpha\alpha'}$,
$(S_\alpha)_{\beta z} (S_\alpha)_{z \beta'} = \delta_{\beta\beta'}$,
$(S_\alpha)_{z \beta} (S_\alpha)_{z \beta'} = (S_\alpha)_{\beta z} (S_\alpha)_{\beta' z} = (\sigma_x)_{\beta\beta'}$,
and $(S_z)_{\beta\beta'} = (\sigma_z)_{\beta\beta'}$, where $\sigma_x$ and $\sigma_z$ are the usual $2 \times 2$ Pauli matrices.
For the transverse Lagrangian density one finds the structure
\begin{widetext}
\begin{equation}
\begin{split}
\mathcal{L}_\perp = {}&{} - \frac{I}{2 \beta_0} \left[ \left( \frac{\dot{I}}{2 I} \right)^2 + \left( \frac{\dot{\vartheta}}{2} \right)^2
+ \dot{\Theta}^2 + \left( \frac{\dot{\chi}}{2} \right)^2 + 2 \cos\vartheta \, \dot{\Theta} \, \frac{\dot{\chi}}{2} \right] \\
{}& \begin{aligned}[t]
{} - \frac{I}{2 \beta_0} \bigg[ {}&{}
(K_{II})_{\alpha\alpha'} \frac{\partial_\alpha I}{2 I} \frac{\partial_{\alpha'} I}{2 I}
+ (K_{\vartheta\vartheta})_{\alpha\alpha'} \frac{\partial_\alpha \vartheta}{2} \frac{\partial_{\alpha'} \vartheta}{2}
+ (K_{\Theta\Theta})_{\alpha\alpha'} \partial_\alpha \Theta \, \partial_{\alpha'} \Theta
+ (K_{\chi\chi})_{\alpha\alpha'} \frac{\partial_\alpha \chi}{2} \frac{\partial_{\alpha'} \chi}{2} \\
{}&{} 
+ (K_{I\vartheta})_{\alpha\alpha'} \frac{\partial_\alpha I}{2 I} \frac{\partial_{\alpha'} \vartheta}{2}
+ (K_{I\Theta})_{\alpha\alpha'} \frac{\partial_\alpha I}{2 I} \, \partial_{\alpha'} \Theta
+ (K_{I\chi})_{\alpha\alpha'} \frac{\partial_\alpha I}{2 I} \, \frac{\partial_{\alpha'} \chi}{2} \\
{}&{}
+ (K_{\vartheta\Theta})_{\alpha\alpha'} \frac{\partial_\alpha \vartheta}{2} \, \partial_{\alpha'} \Theta
+ (K_{\vartheta\chi})_{\alpha\alpha'} \frac{\partial_\alpha \vartheta}{2} \frac{\partial_{\alpha'} \chi}{2}
+ (K_{\Theta\chi})_{\alpha\alpha'} \partial_\alpha \Theta \, \frac{\partial_{\alpha'} \chi}{2} \bigg]
\end{aligned} \\
{}&{} + \frac{\beta_0}{2} I - \frac{g_d}{2} I^2 - \frac{g_s}{2} I^2 \cos^2\vartheta \, ,
\end{split}
\label{eq:dp_lagr_perp}
\end{equation}
while the longitudinal Lagrangian density in the density-phase variables becomes
\begin{equation}
\begin{split}
\mathcal{L}_z = {}&{} - \frac{1}{2 \beta_0}
\left[ \partial_\alpha \hat{\mathcal{E}}_z^* \partial_\alpha \hat{\mathcal{E}}_z + \partial_\alpha \Theta \, \partial_\alpha \Theta |\hat{\mathcal{E}}_z|^2
- \mi (\hat{\mathcal{E}}_z^* \partial_\alpha \hat{\mathcal{E}}_z
- \hat{\mathcal{E}}_z \partial_\alpha \hat{\mathcal{E}}_z^*) \partial_\alpha \Theta \right] \\
{}&{}
\begin{aligned}[t]
- \frac{\sqrt{I}}{\beta_0} \Re \bigg\{ {}&{} \bigg[ (S_\alpha S_z)_{z+}
\left( \cos\frac{\vartheta}{2} \frac{\dot{I}}{2 I} - \sin\frac{\vartheta}{2} \frac{\dot{\vartheta}}{2}
+ \mi \cos\frac{\vartheta}{2} \dot{\Theta} + \mi \cos\frac{\vartheta}{2} \frac{\dot{\chi}}{2} \right) \me^{\mi \chi / 2} \\
{}&{} \phantom{\bigg[} + (S_\alpha S_z)_{z-}
\left( \sin\frac{\vartheta}{2} \frac{\dot{I}}{2 I} + \cos\frac{\vartheta}{2} \frac{\dot{\vartheta}}{2}
+ \mi \sin\frac{\vartheta}{2} \dot{\Theta} - \mi \sin\frac{\vartheta}{2} \frac{\dot{\chi}}{2} \right) \me^{- \mi \chi / 2}
\bigg] (\partial_\alpha \hat{\mathcal{E}}_z^* - \mi \hat{\mathcal{E}}_z^* \partial_\alpha \Theta) \bigg\}
\end{aligned} \\
{}&{} + \frac{\beta_0}{2}| \hat{\mathcal{E}}_z |^2
- g_d I | \hat{\mathcal{E}}_z |^2 - \frac{g_d}{2} | \hat{\mathcal{E}}_z |^4
- g_s I | \hat{\mathcal{E}}_z |^2 - \frac{g_s}{2} I \sin\vartheta (\hat{\mathcal{E}}_z^2 + \hat{\mathcal{E}}_z^{*2}) \, .
\end{split}
\label{eq:dp_lagr_z}
\end{equation}
\end{widetext}
The $K$ coefficients entering $\mathcal{L}_\perp$ are given in Appendix~\ref{app:Kcoeff} for clarity.

\section{Linearized Helmholtz equation}
\label{sec:bogo_theory}
In this section we develop the Bogoliubov theory for a fluid of light described by the Helmholtz equation~\eqref{eq:helm_eq}. It consists
in linearizing this equation about a stationary (with respect to effective-time evolution) background solution, which is given in
Sec.~\ref{subsec:bogo_back}. We achieve this goal by first expanding the Lagrangian density~\eqref{eq:dp_lagr_perp}--\eqref{eq:dp_lagr_z}
up to quadratic order in the fluctuations (Sec.~\ref{subsec:bogo_lag}). Then, we derive their (linear) evolution equations after switching to
the Hamiltonian framework (Sec.~\ref{subsec:bogo_eqs}). Finally, in Sec.~\eqref{subsec:bogo_prop}, we discuss some relevant properties of the
Bogoliubov modes and, in particular, their orthonormalization relations.

\subsection{Background field solution}
\label{subsec:bogo_back}
As mentioned in Sec.~\ref{sec:dens_phase}, the density-phase formalism is well suited for studying the effective-time evolution of the electric-field
fluctuations about a fixed background $\vec{\mathcal{E}}_0$, henceforth assumed to be uniform, i.e., of constant optical intensity $I_0 =
|\vec{\mathcal{E}}_0|^2$. Specifically, from now on, we will focus on the case of a linearly polarized background field describing a plane wave impinging on the
nonlinear medium at normal incidence that subsequently propagates along the positive $z$ direction with wave vector $\vec{k} = k \hat{\vec{e}}_z$.
Without loss of generality, we take the polarization parallel to the $x$ axis and thus write
\begin{equation}
\vec{\mathcal{E}}_0(\vec{r}_\perp,z) = \sqrt{I_0} \, \me^{\mi k z} \hat{\vec{e}}_x \, .
\label{eq:ansatz}
\end{equation}
Notice that $\vec{\mathcal{E}}_0$ can be rewritten in the form~\eqref{eq:dp_var} by taking, for the density-phase variables, the values
$\vartheta_0 = \pi/2$, $\Theta_0 = \pi/2 + k z$, $\chi_0 = \pi$, and $\smash{\hat{\mathcal{E}}_{z,0} = 0}$. Inserting the field~\eqref{eq:ansatz} into the Helmholtz
equation~\eqref{eq:helm_eq} yields the relation
\begin{equation}
k = \sqrt{\beta_0^2 - 2 \beta_0 g_d I_0} \, ,
\label{eq:refr_ind}
\end{equation}
where the second term in the square root can be interpreted as the nonlinear contribution to the refractive index felt by the background field.
Equation~\eqref{eq:ansatz}-\eqref{eq:refr_ind} represents an exact stationary solution of the Helmholtz equation, which remains uniform at
all $z$; thus, unlike nonuniform backgrounds, it is not subject to self-focusing or defocusing phenomena induced by the nonlinearity.
Notice that the input field $\vec{\mathcal{E}}_0(\vec{r}_\perp,z\!=\!0) = \sqrt{I_0} \, \hat{\vec{e}}_x$ does not depend on the nonlinearity,
and is thus continuous at the air-medium interface. On the other hand, its first-order derivative $\dot{\vec{\mathcal{E}}}_0(\vec{r}_\perp,z\!=\!0)
= \mi k \sqrt{I_0} \, \hat{\vec{e}}_x$ depends on the nonlinearity through $k$, which is given by Eq.~\eqref{eq:refr_ind} for $z > 0$, whereas
$k = \beta_0$ for $z < 0$. Note, finally, that Eq.~\eqref{eq:ansatz} implicitly neglects any reflection or shift at the interface, which is a reasonable
approximation as long as the jump in the nonlinearity at the interface is not too large.

\subsection{Bogoliubov Lagrangian}
\label{subsec:bogo_lag}
We are now ready to investigate fluctuations on top of the uniform background field introduced in Sec.~\ref{subsec:bogo_back}. For this purpose,
in the density-phase representation~\eqref{eq:dp_var} we assume that the intensity $I$ and the relative weight between the two transverse
polarization components $\vartheta$ undergo small fluctuations around the background values, i.e., we write $I = I_0 + \delta I$ and $\vartheta = \pi/2
+ \delta\vartheta$, where $|\delta I|/I_0 \ll 1$ and $|\delta\vartheta| \ll 1$. However, such an expansion does not hold for the global phase whose
fluctuations can be significant, as motivated earlier. Nevertheless, it is convenient to redefine it as $\Theta \to \pi/2 + k z + \Theta$ in order to isolate the
background phase. At this stage, let us point out that $\mathcal{L}_\perp$ and $\mathcal{L}_z$ do not depend on $\Theta$ itself but only on its
derivatives. Hence, in order to develop the Bogoliubov theory, it is sufficient to assume the derivatives of $\Theta$ to be small,
which is the typical situation of low-dimensional quantum systems~\cite{Popov1972,Popov_book,Mora2003,Petrov2004}. A different situation occurs
for the relative phase, since the Lagrangian densities~\eqref{eq:dp_lagr_perp} and~\eqref{eq:dp_lagr_z} explicitly depend on $\chi$ because of the
spin-orbit coupling terms. In order to formulate the Bogoliubov approach we hypothesize that such terms suppress the fluctuations of $\chi$
and make the expansion $\chi = \pi + \delta \chi$ with $|\delta\chi| \ll 1$. In this respect, we point out that a phenomenon of suppression of the
relative-phase fluctuations has been found in a model of atomic gases with spin-orbit coupling~\cite{Su2017}. Finally, the emergence of the longitudinal
field $\hat{\mathcal{E}}_z$ is only caused by the fluctuations of the transverse components, such that we expect its magnitude to remain small.

Let us expand the contributions~\eqref{eq:dp_lagr_perp} and~\eqref{eq:dp_lagr_z} of the Lagrangian density
up to second order in the above small variables, $\mathcal{L}_{\perp(z)} = \mathcal{L}^{(0)}_{\perp(z)} + \mathcal{L}^{(1)}_{\perp(z)}
+ \mathcal{L}^{(2)}_{\perp(z)}$. At zero order one finds $\smash{\mathcal{L}_\perp^{(0)} = g_d I_0^2/2}$ and $\smash{\mathcal{L}_z^{(0)} = 0}$. The first-order terms
can be discarded as they have the form of a total divergence, $\mathcal{L}_\perp^{(1)} = - k I_0 \dot{\Theta} / \beta_0$ and $\smash{\mathcal{L}_z^{(1)} =
\sqrt{I_0/2} (k/\beta_0) \partial_\alpha \Re \left\{ \left[ (S_\alpha S_z)_{z+} - (S_\alpha S_z)_{z-} \right] \hat{\mathcal{E}}_z^* \right\}}$. Hence,
we focus our attention on the second-order contributions. It is convenient to introduce the Fourier transforms of the variables with respect to the
transverse plane coordinates. For the total intensity we write $\delta \tilde{I}(\vec{q}_\perp,z) = \int \dif^2 r_\perp \, \delta I(\vec{r}_\perp,z)
\, \me^{- \mi \vec{q}_\perp \cdot \vec{r}_\perp}$, with the property $\delta \tilde{I}(-\vec{q}_\perp,z) = \delta \tilde{I}^*(\vec{q}_\perp,z)$ due to the
reality of $\delta I$. Analogous expressions hold for $\delta\tilde{\vartheta}$, $\tilde{\Theta}$, and $\delta \tilde{\chi}$ (here and in the following, tilded
functions refer to Fourier transforms). Correspondingly, one can define the transverse and longitudinal Lagrangian densities in momentum space, such that
the corresponding total Lagrangians are given by $L_{\perp(z)}^{(2)} = \int \dif^2 r_\perp \, \mathcal{L}_{\perp(z)}^{(2)} = \int \dif^2 q_\perp / (2 \pi)^2 \,
\tilde{\mathcal{L}}_{\perp(z)}^{(2)}$. We first compute the longitudinal Lagrangian density in momentum space:
\begin{widetext}
\begin{equation}
\begin{split}
\tilde{\mathcal{L}}_z^{(2)} = {}&{} - \frac{1}{2\beta_0}
\left[ (q_\perp^2 - k_0^2) \tilde{\mathcal{E}}_z^*(\vec{q}_\perp) \tilde{\mathcal{E}}_z(\vec{q}_\perp)
+ \frac{\Delta k_0^2}{2} \tilde{\mathcal{E}}_z(\vec{q}_\perp) \tilde{\mathcal{E}}_z(-\vec{q}_\perp)
+ \frac{\Delta k_0^2}{2} \tilde{\mathcal{E}}_z^*(\vec{q}_\perp) \tilde{\mathcal{E}}_z^*(-\vec{q}_\perp) \right] \\
&{} - \frac{\sqrt{I_0} \, q_\perp}{2\beta_0} \left[ \mathcal{A}(\vec{q}_\perp) \tilde{\mathcal{E}}_z^*(\vec{q}_\perp)
+ \mathcal{A}^*(\vec{q}_\perp) \tilde{\mathcal{E}}_z(\vec{q}_\perp) \right] \, .
\end{split}
\label{eq:lagr_z_2_Ez}
\end{equation}
\end{widetext}
Here $k_0^2 = \beta_0^2 - 2 \beta_0 (g_n + g_s) I_0$, $\Delta k_0^2 = 2 \beta_0 g_s I_0$, and
\begin{equation}
\begin{split}
\mathcal{A} = {}&{}
\cos\varphi \left( \frac{\delta\dot{\tilde{I}}}{2 I_0} + \mi k \frac{\delta\tilde{I}}{2 I_0} \right)
- \mi \sin\varphi \left( \frac{\delta\dot{\tilde{\vartheta}}}{2} + \mi k \frac{\delta\tilde{\vartheta}}{2} \right) \\
&{} + \mi \cos\varphi \left( \dot{\tilde{\Theta}} + \mi k \tilde{\Theta} \right)
- \sin\varphi \left( \frac{\delta\dot{\tilde{\chi}}}{2} + \mi k \frac{\delta\tilde{\chi}}{2} \right) \, .
\end{split}
\label{eq:lagr_z_2_coeff}
\end{equation}
In writing Eq.~\eqref{eq:lagr_z_2_Ez} we have used the polar form $\vec{q}_\perp = q_\perp (\cos\varphi \, \hat{\vec{e}}_x
+ \sin\varphi \, \hat{\vec{e}}_y)$ of the transverse momentum and the identity $[(\vec{q}_\perp \cdot \vec{S}_\perp) S_z]_{z\pm} =
\pm q_\perp \me^{\pm \mi \varphi} / \sqrt{2}$. Notice that $\tilde{\mathcal{L}}_z$ only contains linear and quadratic terms in $\tilde{\mathcal{E}}_z$,
$\tilde{\mathcal{E}}_z^*$. Hence, the Euler-Lagrange equations $\delta L_z^{(2)}/\delta \tilde{\mathcal{E}}_z^*(\vec{q}_\perp) = 0$ and
$\delta L_z^{(2)}/\delta \tilde{\mathcal{E}}_z(-\vec{q}_\perp) = 0$ are linear in the fields $\tilde{\mathcal{E}}_z(\vec{q}_\perp)$ and
$\tilde{\mathcal{E}}_z^*(-\vec{q}_\perp)$. Solving these equations one gets
\begin{equation}
\tilde{\mathcal{E}}_z(\vec{q}_\perp) = - \frac{\sqrt{I_0} \, q_\perp
\left[(q_\perp^2 - k_0^2) \mathcal{A}(\vec{q}_\perp) - \Delta k_0^2 \mathcal{A}^*(-\vec{q}_\perp) \right]}{(q_\perp^2 - k_0^2)^2 - (\Delta k_0^2)^2} \, .
\label{eq:sol_Ez}
\end{equation}
This result can be used to eliminate $\tilde{\mathcal{E}}_z$ in favor of the other variables of the problem. Inserting it into Eq.~\eqref{eq:lagr_z_2_Ez} one
obtains
\begin{equation}
\begin{split}
\tilde{\mathcal{L}}_z^{(2)} = {}&{} \frac{q_\perp^2 I_0}{4\beta_0}
\frac{(q_\perp^2 - k_0^2) \left[ \left|\mathcal{A}(\vec{q}_\perp)\right|^2 + \left|\mathcal{A}(-\vec{q}_\perp)\right|^2 \right]}
{(q_\perp^2 - k_0^2)^2 - (\Delta k_0^2)^2} \\
&{} -\frac{q_\perp^2 I_0}{2\beta_0}\frac{ \Delta k_0^2 \Re \left[ \mathcal{A}(\vec{q}_\perp) \mathcal{A}(-\vec{q}_\perp)\right]}
{(q_\perp^2 - k_0^2)^2 - (\Delta k_0^2)^2} \, .
\end{split}
\label{eq:lagr_z_2}
\end{equation}
In this expression, we have omitted terms that are odd under exchange of $\vec{q}_\perp$ into $-\vec{q}_\perp$ and thus do not contribute to the Lagrangian
$\smash{L_z^{(2)} = \int {\dif^2 q_\perp}/{(2\pi)^2} \tilde{\mathcal{L}}_z^{(2)}}$. It now remains to substitute Eq.~\eqref{eq:lagr_z_2_coeff} into~\eqref{eq:lagr_z_2}
and rewrite $\smash{\tilde{\mathcal{L}}_z^{(2)}}$ as a function of the four variables $\delta \tilde{I}$, $\delta \tilde{\vartheta}$, $\tilde{\Theta}$, $\delta \tilde{\chi}$,
and their derivatives. Then, one has to compute the transverse Lagrangian density $\smash{\tilde{\mathcal{L}}_\perp^{(2)}}$. The calculation is
straightforward and requires the use of the identities $[(\vec{q}_\perp \cdot \vec{S}_\perp)^2]_{++} = [(\vec{q}_\perp \cdot \vec{S}_\perp)^2]_{--} = q_\perp^2/2$
and $[(\vec{q}_\perp \cdot \vec{S}_\perp)^2]_{+-} = [(\vec{q}_\perp \cdot \vec{S}_\perp)^2]_{-+}^* = q_\perp^2 \me^{- 2 \mi \varphi} /2$. After some algebra one
finally obtains the complete Lagrangian density in momentum space, $\tilde{\mathcal{L}}^{(2)} = \tilde{\mathcal{L}}_\perp^{(2)} + \tilde{\mathcal{L}}_z^{(2)}$. It
exhibits the structure
\begin{equation}
\tilde{\mathcal{L}}^{(2)} = \dot{X}^\dagger \Lambda_2 \dot{X} + \dot{X}^\dagger \Lambda_1 X + X^\dagger \Lambda_1^T \dot{X} - X^\dagger \Lambda_0 X \, ,
\label{eq:dp_lagr_f} 
\end{equation}
where $X = (\delta\tilde{I}/2I_0 \,\,\, \delta\tilde{\vartheta}/2 \,\,\, \tilde{\Theta} \,\,\, \delta\tilde{\chi}/2)^T$ is a four-component column vector and 
$\Lambda_0$, $\Lambda_1$, and $\Lambda_2$ are real $4 \times 4$ matrices. Their expressions are a bit cumbersome and are given in
Appendix~\ref{app:bogo_mat} for clarity.

\subsection{Bogoliubov equations in the Hamiltonian framework}
\label{subsec:bogo_eqs}
To study the evolution of the fluctuations, one could write down the Euler-Lagrange equations associated with the Lagrangian
density~\eqref{eq:dp_lagr_f}. These are second-order linear equations in the effective time $z$, whose solution is unique once the input values of the field
and its derivative, namely $X(\vec{q}_\perp,z=0)$ and $\dot{X}(\vec{q}_\perp,z=0)$, have been specified. In the following, however, we will use a
different approach and work in the Hamiltonian framework, which offers two advantages: the evolution equations are of first order, and a direct procedure
exists for determining the orthonormalization conditions of the Bogoliubov amplitudes.

The starting point of the Hamiltonian scheme consists in introducing the conjugate momenta for each dynamical variable in $X$ and then define a new
four-component column vector collecting them, $\Pi = (\Pi_{\delta \tilde{I}/2I_0} \,\,\, \Pi_{\delta\tilde{\vartheta}/2} \,\,\,
\Pi_{\tilde{\Theta}} \,\,\, \Pi_{\delta\tilde{\chi}/2})^T$. The formal definition of this quantity is $\smash{\Pi = \partial \tilde{\mathcal{L}}^{(2)}/\partial \dot{X}^T}$,
yielding
\begin{equation}
\Pi = \Lambda_2 \dot{X}^* + \Lambda_1 X^* \, .
\label{eq:dp_conj_mom}
\end{equation}
One can now introduce an Hamiltonian density by performing a Legendre transform on the Lagrangian density, $\tilde{\mathcal{H}}^{(2)} = \Pi^T \dot{X}
+ \dot{X}^\dagger \Pi^* - \tilde{\mathcal{L}}^{(2)}$. The derivative of $X$ can be eliminated in favor of $\Pi$ by inverting Eq.~\eqref{eq:dp_conj_mom}.
After some algebra one finds
\begin{equation}
\begin{split}
\tilde{\mathcal{H}}^{(2)} = {}&{}
\Pi^T \Lambda_2^{-1} \Pi^* - \Pi^T (\Lambda_2^{-1} \Lambda_1) X - X^\dagger (\Lambda_2^{-1} \Lambda_1)^T \Pi^* \\
&{} + X^\dagger (\Lambda_1^T \Lambda_2^{-1} \Lambda_1 + \Lambda_0) X \, .
\end{split}
\label{eq:dp_ham_2}
\end{equation}
With this expression at hand, it is straightforward to explicitly derive the Hamilton equations $\dot{X} = \partial \tilde{\mathcal{H}}^{(2)}/\partial \Pi^T$
and $\dot{\Pi}^* = - \partial \tilde{\mathcal{H}}^{(2)}/\partial X^\dagger$ governing the effective-time evolution of $X$ and $\Pi^*$. The final result can
be written in the compact form
\begin{equation}
\mi
\begin{pmatrix}
\dot{X} \\
\dot{\Pi}^*
\end{pmatrix}
= \mathcal{B}
\begin{pmatrix}
X \\
\Pi^*
\end{pmatrix} \, ,
\label{eq:dp_ham_eq}
\end{equation}
where we have defined the $8 \times 8$ Bogoliubov matrix
\begin{equation}
\mathcal{B} = \mi
\begin{pmatrix}
- \Lambda_2^{-1} \Lambda_1 & \Lambda_2^{-1} \\
- (\Lambda_1^T \Lambda_2^{-1} \Lambda_1 + \Lambda_0) & (\Lambda_2^{-1} \Lambda_1)^T
\end{pmatrix} \, .
\label{eq:dp_eig_mat}
\end{equation}
This system of eight linear homogeneous differential equations of first order in the $z$ coordinate admits eight linearly independent solutions.
To identify a set of such solutions we make the Ansatz $X(\vec{q}_\perp,z) = X_0(\vec{q}_\perp) \me^{- \mi \Omega(\vec{q}_\perp) z}$,
$\Pi^*(\vec{q}_\perp,z) = \Pi_0^*(\vec{q}_\perp) \me^{- \mi \Omega(\vec{q}_\perp) z}$, and rewrite Eq.~\eqref{eq:dp_ham_eq} as an eigenvalue problem,
\begin{equation}
\mathcal{B}
\begin{pmatrix}
X_0 \\
\Pi_0^*
\end{pmatrix}
=
\Omega
\begin{pmatrix}
X_0 \\
\Pi_0^*
\end{pmatrix} \, .
\label{eq:dp_eig_eq}
\end{equation}
We will use the subscript $\ell$ to distinguish between different solutions of this problem. The set of eigenfrequencies $\Omega_\ell$ represents
the Bogoliubov spectrum of the system, and the $X_{0,\ell}$'s and $\Pi_{0,\ell}^*$'s are the corresponding amplitudes. Any arbitrary solution of
Eq.~\eqref{eq:dp_ham_eq} can be represented as a linear combination of these Bogoliubov modes,
\begin{equation}
\begin{pmatrix}
X(\vec{q}_\perp,z) \\
\Pi^*(\vec{q}_\perp,z)
\end{pmatrix}
= \sum_{\ell} C_\ell(\vec{q}_\perp)
\begin{pmatrix}
X_{0,\ell}(\vec{q}_\perp) \\
\Pi_{0,\ell}^*(\vec{q}_\perp)
\end{pmatrix}
\me^{- \mi \Omega_\ell(\vec{q}_\perp)z} \, ,
\label{eq:dp_ham_sol}
\end{equation}
where the sum runs over all the modes. The weights $C_\ell(\vec{q}_\perp)$ characterizing this combination are uniquely fixed by the choice of the
input fields $X(\vec{q}_\perp,z\!=\!0)$ and $\Pi^*(\vec{q}_\perp,z=0)$. Their evaluation requires the use of orthonormalization conditions derived in the
next section.

\subsection{Properties of Bogoliubov modes and orthonormalization relations}
\label{subsec:bogo_prop}
An important property of the Bogoliubov formalism is that, if $(X_{0,\ell}(\vec{q}_\perp),\Pi_{0,\ell}^*(\vec{q}_\perp))$ is a solution of the eigenvalue
problem~\eqref{eq:dp_eig_eq} with frequency $\Omega_\ell(\vec{q}_\perp)$, then $(X_{0,\ell}^*(-\vec{q}_\perp),\Pi_{0,\ell}(-\vec{q}_\perp))$ also
is a solution with frequency $-\Omega_\ell^*(-\vec{q}_\perp)$~\cite{Castin_review}.\footnote{The matrix $\mathcal{B}$ and its spectrum are also
symmetric under inversion of $\vec{q}_\perp$ into $- \vec{q}_\perp$. However, this property is lost if the direction of propagation of the background
field is tilted with respect to the $z$ axis. Having in mind this more general case, in the formulas of the present section we keep the explicit distinction
between $\vec{q}_\perp$ and $- \vec{q}_\perp$.} These two solutions correspond to the same physical oscillation of the system. Their simultaneous
appearance is necessary because $X$ and $\Pi^*$ are Fourier transform of real quantities, and must satisfy $X^*(-\vec{q}_\perp,z) = X(\vec{q}_\perp,z)$
and $\Pi(-\vec{q}_\perp,z) = \Pi^*(\vec{q}_\perp,z)$. For the same reason, the weight of the mode of frequency $-\Omega_\ell^*(-\vec{q}_\perp)$ in
the combination~\eqref{eq:dp_ham_sol} must be $C_\ell^*(-\vec{q}_\perp)$ [this comes out automatically when calculating the weights using
Eq.~\eqref{eq:dp_weight}].

We additionally point out that complex frequencies in the spectrum of $\mathcal{B}$ occur in complex conjugate pairs~\cite{Castin_review}. To prove
this, we notice that, if $\Omega_\ell(\vec{q}_\perp)$ is a complex eigenvalue of $\mathcal{B}$, then its conjugate $\Omega_\ell^*
(\vec{q}_\perp)$ must be an eigenvalue of $\mathcal{B}^\dagger$. However, the two matrices are related by a unitary transformation,
\begin{equation}
\begin{pmatrix}
0 & \mi \mathbb{I}_4 \\
- \mi \mathbb{I}_4 & 0
\end{pmatrix}
\mathcal{B}
\begin{pmatrix}
0 & \mi \mathbb{I}_4 \\
- \mi \mathbb{I}_4 & 0
\end{pmatrix}^{-1}
= \mathcal{B}^\dagger \, ,
\label{eq:dp_ham_symm}
\end{equation}
where $\mathbb{I}_4$ denotes the $4 \times 4$ identity matrix. Hence, $\mathcal{B}$ and $\mathcal{B}^\dagger$ must have the same spectrum,
meaning that $\Omega_\ell^*(\vec{q}_\perp)$ is also an eigenvalue of $\mathcal{B}$. It should be kept in mind that complex-frequency modes
correspond to perturbations that either grow or decay exponentially in effective time. In the former case the system rapidly deviates from the
linear regime where the Bogoliubov theory is applicable.

In physical systems whose time evolution is governed by the standard Gross-Pitaevskii equation (including atomic Bose-Einstein condensates and
fluids of light in the paraxial approximation, see Sec.~\ref{sec:paraxial_lim}) the Bogoliubov amplitudes obey a set of orthonormalization
conditions~\cite{Pethick_Smith_book,Pitaevskii_Stringari_book,Castin_review}. We will now prove that this also happens in fluids of light described
by the Helmholtz equation~\eqref{eq:helm_eq}. Writing Eq.~\eqref{eq:dp_eig_eq} for two fixed arbitrary modes $\ell$ and $\ell'$, taking appropriate
combinations of the resulting expressions, and using Eq.~\eqref{eq:dp_ham_symm}, one ends up with the identity
\begin{equation}
\begin{split}
&{} [\Omega_\ell(\vec{q}_\perp)- \Omega_{\ell'}^*(\vec{q}_\perp)] \times \\ 
&{} \mi \left[ X_{0,\ell'}^\dagger(\vec{q}_\perp) \Pi_{0,\ell}^*(\vec{q}_\perp)
- \Pi_{0,\ell'}^T(\vec{q}_\perp) X_{0,\ell}(\vec{q}_\perp) \right] = 0 \, .
\end{split}
\label{eq:dp_herm}
\end{equation}
For modes having real frequency, and in the absence of degeneracy, the second factor on the left-hand side must vanish if $\ell' \neq \ell$. This
provides an orthogonality relation for the amplitudes. The $\ell' = \ell$ case is instead used to define the norm $\mathcal{N}_\ell(\vec{q}_\perp)$
of the $\ell$-th Bogoliubov mode. These conditions summarize as
\begin{equation}
\begin{split}
&{} \mi \left[ X_{0,\ell'}^\dagger(\vec{q}_\perp) \Pi_{0,\ell}^*(\vec{q}_\perp) - \Pi_{0,\ell'}^T(\vec{q}_\perp) X_{0,\ell}(\vec{q}_\perp) \right] \\
&{} = \mathcal{N}_\ell(\vec{q}_\perp) \delta_{\ell' \ell}
\end{split}
\label{eq:dp_norm_r}
\end{equation}
(in this section we do not sum over repeated $\ell$ indices). Notice that the norm is real and can be either positive or negative.
To understand this point, let us assume that $\mathcal{N}_\ell(\vec{q}_\perp) > 0$ for some solution $(X_{0,\ell}(\vec{q}_\perp),
\Pi_{0,\ell}^*(\vec{q}_\perp))$ of Eq.~\eqref{eq:dp_eig_eq}. Then, from Eq.~\eqref{eq:dp_norm_r} one finds that the other solution
$(X_{0,\ell}^*(-\vec{q}_\perp),\Pi_{0,\ell}(-\vec{q}_\perp))$ corresponding to the same physical oscillation has negative norm equal to
$- \mathcal{N}_\ell(-\vec{q}_\perp)$.

The orthonormalization condition of complex-frequency modes is again deduced from Eq.~\eqref{eq:dp_herm} and reads
\begin{equation}
\begin{split}
&{} \mi \left[X_{0,\ell'}^\dagger(\vec{q}_\perp) \Pi_{0,\ell}^*(\vec{q}_\perp) - \Pi_{0,\ell'}^T(\vec{q}_\perp) X_{0,\ell}(\vec{q}_\perp) \right] \\
&{} = \mathcal{N}_\ell(\vec{q}_\perp) \delta_{\ell' \bar{\ell}} \, .
\end{split}
\label{eq:dp_norm_c}
\end{equation}
Here we use the subscript $\bar{\ell}$ to denote the Bogoliubov mode with frequency $\Omega_{\bar{\ell}}(\vec{q}_\perp) =
\Omega_{\ell}^*(\vec{q}_\perp)$. Different from the case of real-frequency modes, here $\mathcal{N}_\ell(\vec{q}_\perp)$ is generally
complex and such that $\mathcal{N}_{\bar{\ell}}(\vec{q}_\perp) = [\mathcal{N}_\ell(\vec{q}_\perp)]^*$.

The orthonormalization conditions of the amplitudes can be used to express the weights in the linear superposition~\eqref{eq:dp_ham_sol}
in terms of the fields at the interface. Indeed, taking Eq.~\eqref{eq:dp_ham_sol} at $z=0$ and projecting it onto the $\ell$-th Bogoliubov mode
using Eq.~\eqref{eq:dp_norm_r} or~\eqref{eq:dp_norm_c} one finds
\begin{equation}
\begin{split}
&{} C_\ell(\vec{q}_\perp) =\frac{\mi}{ \mathcal{N}_{\bar{\ell}}(\vec{q}_\perp)} \times \\
&{} \left[ X_{0,\bar{\ell}}^\dagger(\vec{q}_\perp) \Pi^*(\vec{q}_\perp,z\!=\!0)
\!-\! \Pi_{0,\bar{\ell}}^T(\vec{q}_\perp) X(\vec{q}_\perp,z\!=\!0) \right]
\end{split}
\label{eq:dp_weight}
\end{equation}
(with $\bar{\ell}$ replaced by $\ell$ in the case of a real-frequency mode). Equations~\eqref{eq:dp_eig_eq} and~\eqref{eq:dp_ham_sol}, together
with the relation~\eqref{eq:dp_weight}, constitute the complete solution of the linearized problem.

\section{Paraxial limit}
\label{sec:paraxial_lim}
While the formalism developed so far is exact, in many practical situations one can obtain a simplified description by performing the so-called paraxial
approximation (see, e.g., Refs.~\cite{Landau_Lifshitz_08_book,Agrawal_book}). In this section we recall the main elements of this approximation, and
show how it relates to the more general framework developed in the previous sections. In particular, we show that taking into account the polarization
degrees of freedom in the paraxial approximation establishes a mapping between the fluid of light and a binary Bose mixture described by two coupled
Gross-Pitaevskii equations.

\subsection{Paraxial approximation and Gross-Pitaevskii theory}
\label{subsec:paraxial_gp}
Let us consider a light beam mostly propagating along the $z$ axis, and write the complex electric field in the form
\begin{equation}
\vec{\mathcal{E}}(\vec{r}_\perp,z) = \vec{\psi}(\vec{r}_\perp,z) \me^{\mi \beta_0 z} \, .
\label{eq:paraxial_env}
\end{equation}
Here $\me^{\mi \beta_0 z}$ is a rapidly varying exponential and $\vec{\psi}(\vec{r}_\perp,z)$ a slowly varying function of $\vec{r}_\perp$ and $z$.
Taking the divergence on both sides of Eq.~\eqref{eq:helm_eq} and of its complex conjugate, one finds after a few manipulations
\begin{equation}
\nabla \cdot \vec{\mathcal{E}} =
\frac{\left(\beta_0 - 2 g_I \left| \vec{\psi} \right|^2 \right)\mathcal{G} + 2 g_P \vec{\psi}^2 \me^{2 \mi \beta_0 z} \mathcal{G}^*}
{\left(\beta_0 - 2 g_I \left| \vec{\psi} \right|^2 \right)^2 - \left| 2 g_P \vec{\psi}^2 \right|^2} \, ,
\label{eq:paraxial_field_div}
\end{equation}
where in the right-hand side we replaced $\vec{\mathcal{E}}$ by $\vec{\psi}$ using Eq.~\eqref{eq:paraxial_env} and we defined
\begin{equation}
\mathcal{G} = \vec{\psi} \cdot \nabla \left(2 g_I \left| \vec{\psi} \right|^2\right) \me^{\mi \beta_0 z}
+ \vec{\psi}^* \cdot \nabla \left(2 g_P \vec{\psi}^2 \me^{2 \mi \beta_0 z}\right) \me^{- \mi \beta_0 z} \, .
\label{eq:paraxial_g}
\end{equation}
In typical situations, the quantity $\nabla \cdot \vec{\mathcal{E}}$ is small for two reasons. First, because the nonlinear terms are weak,
$|g_{I(P)}| \left| \vec{\psi} \right|^2 \ll \beta_0$. Second, because it is proportional to the derivatives of the slowly varying envelop $\vec{\psi}$. 
We further notice that in these conditions one can approximate $|\nabla \cdot \vec{\mathcal{E}}| \simeq \beta_0 |\psi_z|$, which implies
that the longitudinal component of the electric field remains small inside the medium, $\left| \psi_z \right| \ll \left| \psi_\pm \right|$.

The paraxial approximation amounts to completely neglecting contributions due to the longitudinal component $\psi_z$, as well as those associated
with the second term of Eq.~\eqref{eq:helm_eq}. As the latter couples spatial and polarization degrees of freedom, it follows that any effect of spin-orbit
coupling is discarded at the paraxial level. By inserting Eq.~\eqref{eq:paraxial_env} into~\eqref{eq:helm_lagr} and applying the paraxial approximation,
we obtain the paraxial Lagrangian density:
\begin{equation}
\begin{split}
\mathcal{L}_\mathrm{par} = {}&{} \frac{\mi}{2} \left( \psi_\beta^* \dot{\psi}_\beta - \psi_\beta \dot{\psi}_\beta^* \right)
- \frac{\partial_\alpha \psi_\beta^* \, \partial_\alpha \psi_\beta}{2 \beta_0} \\
{}&{} - \frac{g_d}{2} \left( |\psi_+|^2 + |\psi_-|^2 \right)^2 - \frac{g_s}{2} \left( |\psi_+|^2 - |\psi_-|^2 \right)^2 \, .
\end{split}
\label{eq:paraxial_lagr}
\end{equation}
This Lagrangian density is formally identical to that of a two-dimensional Bose-Bose mixture~\cite{Pethick_Smith_book,Pitaevskii_Stringari_book}.
The parameter $\beta_0$ plays the role of the atom mass, which is the same for the two components. The two intracomponent couplings are equal
and given by $g_d + g_s$, while the strength of the intercomponent interaction is $g_d - g_s$. The Euler-Lagrange equations for $\psi_\pm$ derived
from the Lagrangian density~\eqref{eq:paraxial_lagr} have the standard form of two coupled nonlinear Schr\"{o}dinger (or Gross-Pitaevskii) equations
for the two components of circularly polarized light:
\begin{equation}
\mi \dot{\psi}_\pm = - \frac{\nabla_\perp^2 \psi_\pm}{2 \beta_0}
+ \left[ (g_d + g_s) \left| \psi_\pm \right|^2 + \left(g_d - g_s\right) \left| \psi_\mp \right|^2 \right] \psi_\pm \, .
\label{eq:paraxial_eqs}
\end{equation}
In the optical context, a further simplification is obtained when the electric field is linearly polarized at any point in space, that is, $\psi_+ =
\me^{\mi \theta} \psi_- \equiv \psi / \sqrt{2}$ for arbitrary $\theta$. In this case the coupled equations~\eqref{eq:paraxial_eqs} reduce to a single
nonlinear Schr\"{o}dinger equation for the wave function $\psi$,
\begin{equation}
\mi \dot{\psi} = - \frac{\nabla_\perp^2 \psi}{2 \beta_0} + g \left| \psi \right|^2 \psi
\label{eq:paraxial_gp}
\end{equation}
with $g = g_d$. The same holds if the field is circularly polarized, i.e., $\psi_+ \equiv \psi$, $\psi_- = 0$ (or $\psi_+ = 0$, $\psi_- \equiv \psi$),
but with a different nonlinear coupling $g = g_d + g_s$. These observations suggest an interesting method to access the values of $g_d$ and $g_s$,
by comparing measurements of the coupling strength $g$ of Eq.~\eqref{eq:paraxial_gp} performed using linearly and circularly polarized light.

The density-phase formalism introduced in Sec.~\ref{sec:dens_phase} can be used within the paraxial approximation as well. The parametrization
of the fields $\psi_\pm$,
\begin{equation}
\begin{pmatrix}
\psi_+ \\
\psi_-
\end{pmatrix}
=
\sqrt{I} \, \me^{\mi \Theta}
\begin{pmatrix}
\cos \frac{\vartheta}{2} \, \me^{\mi \chi / 2} \\
\sin \frac{\vartheta}{2} \, \me^{- \mi \chi / 2}
\end{pmatrix},
\label{eq:paraxial_dp_var}
\end{equation}
is analogous to that of $\mathcal{E}_\pm$, see Eq.~\eqref{eq:dp_var}, except that the term $\beta_0 z$ is subtracted from the global phase
$\Theta$ because of the definition~\eqref{eq:paraxial_env}. The Lagrangian density~\eqref{eq:paraxial_lagr} as a function of the density and phase
variables reads
\begin{equation}
\begin{split}
\mathcal{L}_\mathrm{par} = {}&{}
- I \dot{\Theta} - I \cos\vartheta \frac{\dot{\chi}}{2} - \frac{g_d}{2} I^2 - \frac{g_s}{2} I^2 \cos^2\vartheta \\
&{} - \frac{I}{2 \beta_0} \bigg( \left|\frac{\nabla_\perp I}{2 I}\right|^2 + \left|\frac{\nabla_\perp \vartheta}{2}\right|^2
+ \left|\nabla_\perp \Theta\right|^2 + \left|\frac{\nabla_\perp \chi}{2}\right|^2 \\
&{} + 2 \cos\vartheta \, \nabla_\perp \Theta \cdot \frac{\nabla_\perp \chi}{2} \bigg) \, .
\end{split}
\label{eq:paraxial_dp_lagr}
\end{equation}

\subsection{Bogoliubov theory in the paraxial regime}
\label{subsec:paraxial_bogo}
The Bogoliubov theory for a fluid of light in the paraxial limit mirrors that of a two-component Bose mixture, a topic which has already been extensively explored (see,
e.g., the books~\cite{Pethick_Smith_book,Pitaevskii_Stringari_book} and references therein). Let us take again a background field propagating
along $z$ and with linear polarization parallel to the $x$ axis. Its expression is readily found by solving Eq.~\eqref{eq:paraxial_eqs}:
\begin{equation}
\psi_\pm(\vec{r}_\perp,z) = \mp \sqrt{\frac{I_0}{2}} \, \me^{- \mi g_d I_0 z} \, .
\label{eq:paraxial_ansatz}
\end{equation}
It could be equally obtained from the beyond-paraxial result~\eqref{eq:ansatz} in the limit of weak nonlinearity, $|g_d| I_0 / \beta_0 \ll 1$, in which
one can approximate $k \approx \beta_0 - g_d I_0$. Notice that this configuration mimics the behavior of a balanced binary mixture of bosonic atoms.

As in Sec.~\ref{sec:bogo_theory}, we consider small fluctuations of the optical intensities of the two polarization components about the background
field and write $I = I_0 + \delta I$ and $\vartheta = \pi/2 + \delta\vartheta$. We also redefine the phases as $\Theta \to \pi / 2 - g_d I_0 z + \Theta$
and $\chi \to \pi + \chi$, and we assume that the derivatives of the redefined variables remain small [here $\chi$ is not expanded because the paraxial
Lagrangian~\eqref{eq:paraxial_dp_lagr} depends on its derivatives only]. This enables us to expand the Lagrangian density~\eqref{eq:paraxial_lagr}
up to second order in the small fluctuations, $\mathcal{L}_\mathrm{par} = \mathcal{L}_\mathrm{par}^{(0)} + \mathcal{L}_\mathrm{par}^{(1)}
+ \mathcal{L}_\mathrm{par}^{(2)}$. As usual, the zero-order term $\mathcal{L}_\mathrm{par}^{(0)} = g_d I_0^2 / 2$ is a constant and the first-order one
$\mathcal{L}_\mathrm{par}^{(1)} = - I_0 \dot{\Theta}$ a total divergence. Regarding the second-order contribution, we follow the same procedure
as in Sec.~\ref{subsec:bogo_lag} and work in momentum space. The resulting Lagrangian density can be put in the compact form
\begin{equation}
\tilde{\mathcal{L}}_\mathrm{par}^{(2)}
= \dot{X}^\dagger \Lambda_{\mathrm{par},1} X + X^\dagger \Lambda_{\mathrm{par},1}^T \dot{X} - X^\dagger \Lambda_{\mathrm{par},0} X \, .
\label{eq:paraxial_dp_lagr_f} 
\end{equation}
Here $X = (\delta\tilde{I}/2I_0 \,\,\, \delta\tilde{\vartheta}/2 \,\,\, \tilde{\Theta} \,\,\, \tilde{\chi}/2)^T$ (notice that it differs from the object defined in
Sec.~\ref{subsec:bogo_lag} because one has $\tilde{\chi}$ instead of $\delta\tilde{\chi}$ in the fourth component), and we have introduced the two
$4 \times 4$ matrices
\begin{equation}
\Lambda_{\mathrm{par},1} = I_0
\begin{pmatrix}
0 & 0 & 0 & 0 \\
0 & 0 & 0 & 0 \\
-1 & 0 & 0 & 0 \\
0 & 1 & 0 & 0 \\
\end{pmatrix}
\label{eq:paraxial_dp_lagr_mat_1}
\end{equation}
and
\begin{equation}
\Lambda_{\mathrm{par},0} = I_0 \operatorname{diag}
\left( \frac{q_\perp^2}{2\beta_0} + 2 g_d I_0, \frac{q_\perp^2}{2\beta_0} + 2 g_s I_0, \frac{q_\perp^2}{2\beta_0}, \frac{q_\perp^2}{2\beta_0} \right) \, .
\label{eq:paraxial_dp_lagr_mat_0}
\end{equation}
Alternatively, the paraxial Lagrangian density~\eqref{eq:paraxial_dp_lagr_f} can be derived from Eq.~\eqref{eq:dp_lagr_f}
by expanding the latter up to first order in the dimensionless quantities $\dot{X}/\beta_0$, $(q_\perp/\beta_0)^2$, and $g_{d,s}I_0 / \beta_0$,
which are small in the paraxial limit.

Since the Lagrangian density~\eqref{eq:paraxial_dp_lagr_f} is now of first order in the effective-time derivatives of $X$, unlike in Sec.~\ref{subsec:bogo_eqs}
it is here more convenient to work in the Lagrangian framework. The effective-time evolution is governed by the
Euler-Lagrange equations $\dif(\partial \tilde{\mathcal{L}}_\mathrm{par}^{(2)}/\partial \dot{X}^\dagger)/\dif z 
- \partial \tilde{\mathcal{L}}_\mathrm{par}^{(2)}/\partial X^\dagger = 0$, that is,
\begin{equation}
(\Lambda_{\mathrm{par},1} - \Lambda_{\mathrm{par},1}^T) \dot{X} + \Lambda_{\mathrm{par},0} X = 0 \, .
\label{eq:paraxial_dp_el_eq}
\end{equation}
As compared to Eq.~\eqref{eq:dp_ham_eq}, here we have four coupled equations instead of eight leading to four linearly independent solutions.
They can be found making the Ansatz $X(\vec{q}_\perp,z) = X_0(\vec{q}_\perp) \me^{- \mi \Omega(\vec{q}_\perp) z}$ and rewriting
Eq.~\eqref{eq:paraxial_dp_el_eq} as an eigenvalue problem, $\mathcal{B}_\mathrm{par} X_0 = \Omega X_0$ with
\begin{equation}
\mathcal{B}_\mathrm{par} = \mi (\Lambda_{\mathrm{par},1}^T - \Lambda_{\mathrm{par},1})^{-1} \Lambda_{\mathrm{par},0} \, .
\label{eq:paraxial_dp_eig_mat}
\end{equation}

Most of the properties of the Bogoliubov modes discussed in Sec.~\ref{subsec:bogo_eqs} also hold in the paraxial description. In particular,
for the real-frequency modes, the orthonormalization conditions for the amplitudes read
\begin{equation}
\mi X_{0,\ell'}^\dagger(\vec{q}_\perp) (\Lambda_{\mathrm{par},1} - \Lambda_{\mathrm{par},1}^T) X_{0,\ell}(\vec{q}_\perp)
= \pm \frac{\delta_{\ell \ell'}}{2} \, .
\label{eq:paraxial_dp_norm_r}
\end{equation}
Here the value of the norm of $X_{0,\ell}(\vec{q}_\perp)$ has been chosen equal to $\pm 1/2$ to be consistent with the standard convention adopted
in the study of atomic Bose-Einstein condensates~\cite{Pethick_Smith_book,Pitaevskii_Stringari_book}. Then, if one represents a given solution of
Eq.~\eqref{eq:paraxial_dp_el_eq} as a linear superposition of the kind
\begin{equation}
X(\vec{q}_\perp,z) = \sum_{\ell} C_\ell(\vec{q}_\perp) X_{0,\ell}(\vec{q}_\perp) \me^{- \mi \Omega_\ell(\vec{q}_\perp)z} \, ,
\label{eq:paraxial_dp_el_sol}
\end{equation}
the weights are related to the input value $X(\vec{q}_\perp,z=0)$ by
\begin{equation}
C_\ell(\vec{q}_\perp) = \pm 2\mi X_{0,\ell}^\dagger(\vec{q}_\perp)(\Lambda_{\mathrm{par},1}-\Lambda_{\mathrm{par},1}^T)X(\vec{q}_\perp,z=0) \, .
\label{eq:paraxial_dp_weight_r}
\end{equation}

The diagonalization of $\mathcal{B}_\mathrm{par}$ yields four eigenfrequencies, denoted by $\pm \Omega_d$ and $\pm \Omega_s$. The subscripts
$d$ and $s$ here stand for ``density'' and ``spin'', in a sense that will be clarified below. The eigenfrequencies have the well-known Bogoliubov-like form
\begin{equation}
\Omega_{d(s)}(\vec{q}_\perp) = \sqrt{\frac{q_\perp^2}{2 \beta_0} \left( \frac{q_\perp^2}{2 \beta_0} + 2 \beta_0 c_{d(s)}^2\right)} \, ,
\label{eq:paraxial_spectrum}
\end{equation}
where $\smash{c_{d(s)}^2 = g_{d(s)} I_0 / \beta_0}$ are the two sound velocities. In fluids of light, this dispersion relation was recently measured
in~\cite{Fontaine2018,Stepanov2019}.

The nature of density and spin modes can be understood by looking at the eigenvectors of $\mathcal{B}_\mathrm{par}$. The normalized amplitudes
associated with the positive eigenfrequencies are given by
\begin{subequations}
\label{eq:paraxial_norm_amp}
\begin{align}
X_{0,d}(\vec{q}_\perp) &{} =
\begin{pmatrix}
\displaystyle \frac{1}{2} \sqrt{\frac{q_\perp^2/2 \beta_0}{\Omega_d(\vec{q}_\perp)}}
& 0
& \displaystyle \frac{1}{2\mi} \sqrt{\frac{\Omega_d(\vec{q}_\perp)}{q_\perp^2/2 \beta_0}}
& 0
\end{pmatrix}^T \, ,
\label{eq:paraxial_norm_amp_d} \\
X_{0,s}(\vec{q}_\perp) &{} =
\begin{pmatrix}
0
& \displaystyle \frac{1}{2} \sqrt{\frac{q_\perp^2/2 \beta_0}{\Omega_s(\vec{q}_\perp)}}
& 0
&{} \displaystyle - \frac{1}{2\mi} \sqrt{\frac{\Omega_s(\vec{q}_\perp)}{q_\perp^2/2 \beta_0}}
\end{pmatrix}^T \, .
\label{eq:paraxial_norm_amp_s}
\end{align}
\end{subequations}
These expressions correspond to $\smash{\delta\tilde{\vartheta} / 2 = \tilde{\chi} / 2 = 0}$ for the $d$ modes, and $\smash{\delta\tilde{I} / 2 I_0
= \tilde{\Theta} = 0}$ for the $s$ modes. Hence, the $d$ modes are indeed of pure density type, in the sense that only the total optical intensity $I$ and phase
$\Theta$ oscillate with $z$. In contrast, the $s$ modes are of pure spin type, featuring
oscillations of the relative optical intensity $|\mathcal{E}_+|^2 - |\mathcal{E}_-|^2 = - I_0 \delta\vartheta$ and phase $\chi$.

\section{Exact Bogoliubov spectrum}
\label{sec:bogo_spectrum}
We now examine the Bogoliubov spectrum \emph{without} using the paraxial approximation. This spectrum is found by solving
the exact eigenvalue problem~\eqref{eq:dp_eig_eq}. It includes eight branches, corresponding to four different physical oscillations of the system.
The frequencies are easy to compute in the absence of nonlinearities ($g_d = g_s = 0$). In this limit each Bogoliubov mode is twofold degenerate,
thus one has only four distinct eigenfrequencies, $\pm \Omega_+(\vec{q}_\perp)$ and $\pm \Omega_-(\vec{q}_\perp)$ with $\Omega_\pm(\vec{q}_\perp)
= \beta_0 \pm \sqrt{\beta_0^2 - q_\perp^2}$. This spectrum can be represented by two circumferences in the $(q_\perp,\Omega)$ plane,
having centers $(0,\pm\beta_0)$ and radius $\beta_0$. Consequently, the frequencies are real for $q_\perp \leq \beta_0$, whereas they
become complex when $q_\perp > \beta_0$. At small $q_\perp$, the lowest positive-frequency branch exhibits the standard quadratic
dispersion $\Omega_-(\vec{q}_\perp) \simeq q_\perp^2 / 2 \beta_0$, whereas the upper one is characterized by a gap $\Omega_+(0) =
2 \beta_0$. In terms of the solutions of the original problem [the Helmholtz equation~\eqref{eq:helm_eq} without the nonlinear part]
the lower branch is associated with transmitted modes of the electric field $\vec{\mathcal{E}}$ whose wave vector $\vec{q} =
(\vec{q}_\perp,q_z)$ has positive component along $z$, $q_z = \beta_0 - \Omega_- = \sqrt{\beta_0^2 - q_\perp^2}$. Conversely, modes
belonging to the upper branch have negative $z$ component $\beta_0 - \Omega_+ = - q_z$ of the wave vector, and represent the reflected
part of the field. This branch does not show up in the paraxial description of Sec.~\ref{subsec:paraxial_bogo}, which requires slow effective-time
variations of the field fluctuations (see Sec.~\ref{sec:paraxial_lim}). On the other hand, gapped excitations occur in relativistic Bose-Einstein
condensates, for which they describe the phenomenon of creation of particle-antiparticle pairs~\cite{Fagnocchi2010}.

The twofold degeneracy of the noninteracting Bogoliubov spectrum is lifted when taking nonlinearities into account. One still finds four
solutions with lower frequency, $\pm \Omega_{-,d}(\vec{q}_\perp)$ and $\pm \Omega_{-,s}(\vec{q}_\perp)$, and four with higher frequency,
$\pm\Omega_{+,d}(\vec{q}_\perp)$ and $\pm\Omega_{+,s}(\vec{q}_\perp)$. The lower-frequency solutions reduce to the paraxial
spectra~\eqref{eq:paraxial_spectrum} in the limit $|g_{d,s}| I_0 / \beta_0 \ll 1$ and $(q_\perp/\beta_0)^2 \ll 1$. At small $q_\perp$, the lower
density branch is characterized by the phononlike dispersion $\Omega_{-,d}(\vec{q}_\perp) \simeq c_d q_\perp$, with an isotropic sound velocity
\begin{equation}
c_d^2 = \frac{g_d I_0}{\beta_0 - 3 g_d I_0} \, .
\label{eq:dp_sound_vel_d}
\end{equation}
As in the paraxial limit, the existence of this phonon mode is due to the fact that the background field~\eqref{eq:ansatz} spontaneously breaks
the invariance of the Helmholtz Lagrangian~\eqref{eq:helm_lagr} under global $\mathrm{U}(1)$ transformations of the kind $\vec{\mathcal{E}} \to
\me^{\mi \theta} \vec{\mathcal{E}}$. A similar phenomenon happens for the lower spin branch, whose low-$q_\perp$ behavior $\Omega_{-,s}(\vec{q}_\perp)
\simeq c_s(\varphi) q_\perp$ is linear as well, with a sound velocity
\begin{equation}
c_s^2(\varphi) = \frac{g_s I_0 [\beta_0 - 2 (g_d + 2 g_s \cos^2\varphi) I_0]}{[\beta_0 - (2 g_d + g_s) I_0][\beta_0 - 2 (g_d + 2 g_s) I_0]} \, ,
\label{eq:dp_sound_vel_s}
\end{equation}
where we recall that $\varphi$ is the angle between $\vec{q}_\perp$ and the $x$ axis. This second phonon mode arises because our linearly
polarized background spontaneously breaks the rotational symmetry about the $z$ axis exhibited by the Helmholtz Lagrangian~\eqref{eq:helm_lagr}. 
As compared with the noninteracting limit, we observe that the spin sound velocity~\eqref{eq:dp_sound_vel_s} is \emph{anisotropic}, i.e. it depends
on the direction of the wave vector $\vec{q}_\perp$ relative to the background polarization vector $\hat{\vec{e}}_x$. This property is also a major
difference compared to the paraxial approximation, Eq.~\eqref{eq:paraxial_spectrum}, and a direct signature of spin-orbit coupling of light. A similar feature is
shared by several models of atomic Bose gases with spin-orbit coupling~\cite{Martone2012,Liao2013}. Note that in this work we assume that both $c_d^2$
and $c_s^2$ are positive, which requires $g_d$ and $g_s$ to be themselves positive and not too large. 

\begin{figure}[htb]
\includegraphics[scale=1]{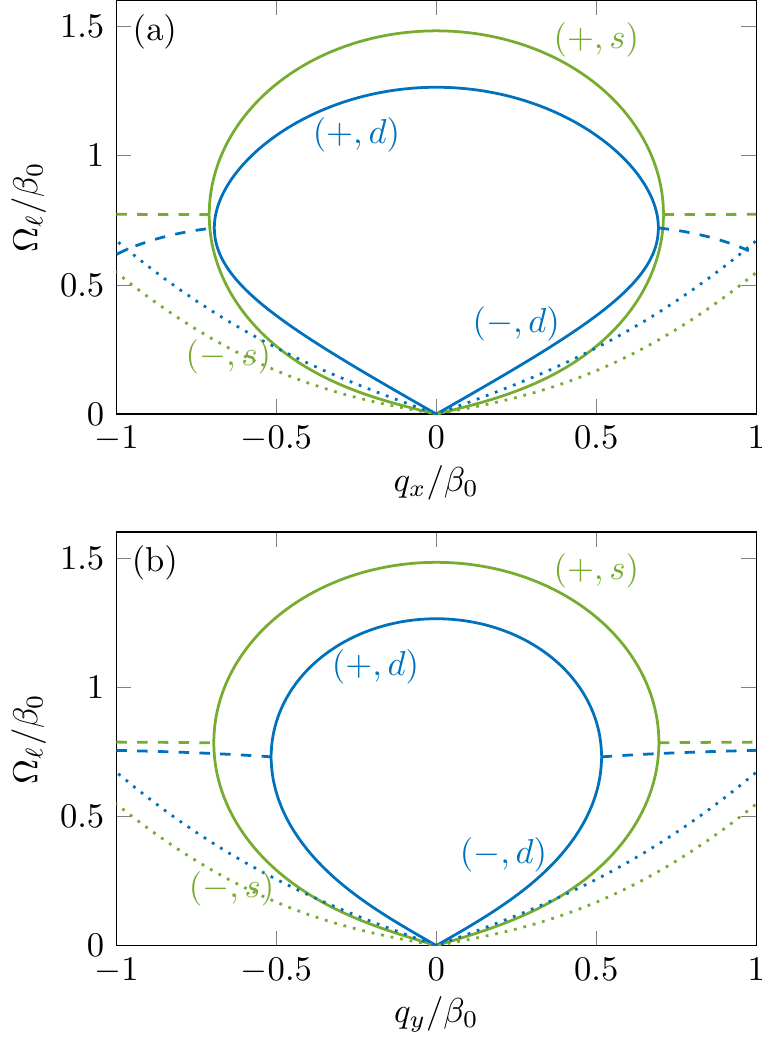}
\caption{Bogoliubov spectrum of a linearly polarized fluid of light as a function of the excitation momentum. The upper (lower) panel reports the result for
$\vec{q}_\perp$ oriented along the $x$ ($y$) axis, that is, parallel (perpendicular) to the polarization direction of the background field. Notice the anisotropy
of the dispersion. The blue (green) curves correspond to density, $\ell=(\pm,d)$ [spin, $\ell=(\pm,s)$] branches, as also indicated in their labels. The solid parts
of these lines (small $q_\perp$) identify modes having real frequency, while the dashed parts (large $q_\perp$) are associated with complex-frequency modes;
in the latter case only the real part of $\Omega_\ell$ is shown. For the sake of comparison we also plot the predictions~\eqref{eq:paraxial_spectrum} of the
paraxial approximation (dotted curves). The nonlinear coupling strengths are $g_d I_0 / \beta_0 = 0.2$ and $g_s I_0 / \beta_0 = 0.05$.}
\label{fig:dp_bogo_sp}
\end{figure}

The dispersion relations $\Omega_{\pm,d}$ and $\Omega_{\pm,s}$ are displayed in Fig.~\ref{fig:dp_bogo_sp} for a fixed choice of the nonlinear coupling
strengths (we do not show their explicit analytic expressions, which are very cumbersome at arbitrary $q_\perp$ and $\varphi$).\footnote{The only
exception is represented by the spin modes for $g_s = 0$, which keep the same circular shape as in the noninteracting case, only with a different radius:
$\Omega_{\pm,s}(\vec{q}_\perp) = k \pm \sqrt{k^2 - q_\perp^2}$ where $k$ is given by Eq.~\eqref{eq:refr_ind}.} At arbitrary values of $q_\perp$, both
dispersions are anisotropic. To illustrate this property, we plot them for $\vec{q}_\perp$ oriented along (upper panel) and perpendicular (lower panel)
to the $x$ axis. Anisotropy of the dispersion relation is again an important difference from the paraxial limit, already visible at low $q_\perp$ in the lower
branches, see Fig.~\ref{fig:dp_bogo_sp}. Note that the frequencies are real up to a critical value of the transverse momentum. Once this value is exceeded
they turn complex and such that $\Omega_{+,d(s)} = [\Omega_{-,d(s)}]^*$. This is a manifestation of the phenomenon of total internal reflection, occurring
in light beams with large incident angle against the interface between two media, which are totally reflected back into the first medium. The critical transverse
momentum coincides with $\beta_0$ in the linear problem, and becomes smaller in the presence of positive nonlinear couplings $g_d$ and $g_s$; besides,
it is different for the $d$ and $s$ modes and generally depends on $\varphi$.

Concerning the upper branches, finally, they remain gapped but the value of the gap is changed by the nonlinearity,
$\Omega_{+,d}(0) = 2 \sqrt{\beta_0^2 - 3 \beta_0 g_d I_0}$ and $\Omega_{+,s}(0) = 2 \sqrt{\beta_0^2 - \beta_0 (2 g_d + g_s) I_0}$.

We complete this section by briefly commenting on the amplitudes of these Bogoliubov modes, focusing on the regime where their frequencies are real.
The amplitude vectors have the structure
\begin{subequations}
\label{eq:dp_norm_amp_X}
\begin{align}
X_{0,\pm,d} &{} =
\begin{pmatrix}
\displaystyle \frac{\delta\tilde{I}_{\pm,d}}{2 I_0}
& \displaystyle \mi \frac{\delta\tilde{\vartheta}_{\pm,d}}{2}
& \mi \tilde{\Theta}_{\pm,d}
& \displaystyle \frac{\delta\tilde{\chi}_{\pm,d}}{2}
\end{pmatrix}^T \, ,
\label{eq:dp_norm_amp_X_d} \\
X_{0,\pm,s} &{} =
\begin{pmatrix}
\displaystyle \mi \frac{\delta\tilde{I}_{\pm,s}}{2 I_0}
& \displaystyle \frac{\delta\tilde{\vartheta}_{\pm,s}}{2}
& \tilde{\Theta}_{\pm,s}
& \displaystyle \mi \frac{\delta\tilde{\chi}_{\pm,s}}{2}
\end{pmatrix}^T \, .
\label{eq:dp_norm_amp_X_s}
\end{align}
\end{subequations}
Here the $\delta\tilde{I}_\ell$'s, $\delta\tilde{\vartheta}_\ell$'s, $\tilde{\Theta}_\ell$'s, and $\delta\tilde{\chi}_\ell$'s are real functions
of $\vec{q}_\perp$. While we have not been able to derive a simple analytical expression for these functions, in practice they can be easily
computed by numerically solving the eigenvalue problem~\eqref{eq:dp_eig_eq}. We find that when $\vec{q}_\perp$ is
parallel or orthogonal to the polarization direction (the $x$ axis) one has $\delta\tilde{\vartheta}_{\pm,d} / 2 = \delta\tilde{\chi}_{\pm,d} / 2 = 0$
and $\delta\tilde{I}_{\pm,s} / 2 I_0 = \tilde{\Theta}_{\pm,s} = 0$. Hence, in these situations the $d$'s are pure density modes, while the $s$'s
are pure spin modes. It turns out, however, that for intermediate directions of $\vec{q}_\perp$ these modes generally exhibit an hybrid
density and spin character, with both the total and relative intensity oscillating simultaneously. This hybridization phenomenon will
be discussed in the next section.

\section{Spin-orbit mode hybridization}
\label{sec:hybr_mode}
\begin{figure}
\includegraphics[scale=0.8]{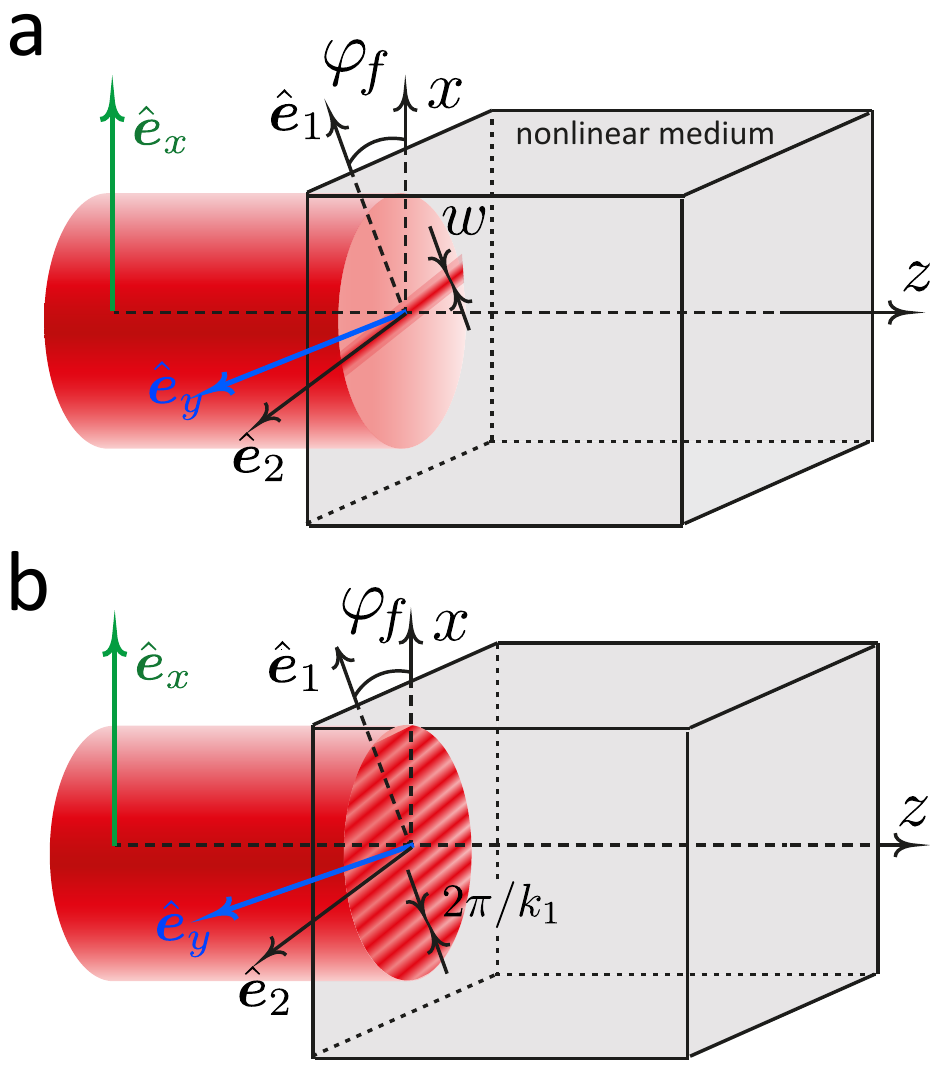}
\caption{Schematic representation of the two experimental scenarios considered in Sec.~\ref{sec:hybr_mode}, involving (a) a small Gaussian probe
and (b) a pure phase perturbation on top of a homogeneous background field sent through a bulk nonlinear medium.}
\label{fig:hybr_configs}
\end{figure}

In this section, we apply the Bogoliubov formalism developed in the previous sections to unveil a phenomenon of mode hybridization in a
spin-orbit-coupled fluid of light. To this end, we consider a concrete experimental scenario where a small probe beam is sent through a
nonlinear material driven by a homogeneous background field. Such a strategy was recently used to experimentally measure the Bogoliubov
dispersion in a (single-component) fluid of light~\cite{Fontaine2018}. Here we address both the cases where the probe beam is a Gaussian
wave packet, Sec.~\ref{subsec:hybr_gauss_exp}, and a pure phase perturbation, Sec.~\ref{subsec:hybr_phase_exp}. A schematic representation
of the two situations is given in Fig.~\ref{fig:hybr_configs}. In both protocols we choose the field profile at the air-medium interface and we study
its evolution inside the nonlinear medium. We do the calculations both in the paraxial regime and slightly beyond it, to show evidence for the
mechanism of mode hybridization. To this end, we focus on small-momentum excitations that populate only the low-frequency (weakly nonparaxial)
modes of the Bogoliubov spectrum. From a theoretical point of view, the latter condition can be fulfilled by appropriately choosing the conjugate
momenta at the interface, as detailed in Appendix~\ref{app:conj_mom_int}.

\subsection{Small Gaussian probe experiment}
\label{subsec:hybr_gauss_exp}
We first consider as the input field a small, Gaussian-shaped wave packet fluctuation on top of a linearly polarized background, as illustrated
in Fig.~\ref{fig:hybr_configs}(a). More specifically, the fluctuation has a purely Gaussian profile along a given direction $\hat{\vec{e}}_1 =
(\cos\varphi_f,\sin\varphi_f)$ in the $(x,y)$ plane, and it is flat in the orthogonal direction $\hat{\vec{e}}_2$. The total field is (recall that we only
need to specify the transverse components)
\begin{equation}
\begin{split}
&{} \begin{pmatrix}
\mathcal{E}_+(\vec{r}_\perp,z=0) \\
\mathcal{E}_-(\vec{r}_\perp,z=0)
\end{pmatrix}= \\
&{} \sqrt{I_0} \left[
\begin{pmatrix}
- 1 / \sqrt{2} \\
1 / \sqrt{2}
\end{pmatrix}
+ \epsilon
\begin{pmatrix}
- \cos\frac{\vartheta_f}{2} \\
\sin\frac{\vartheta_f}{2}
\end{pmatrix}
\exp\left(- \frac{x_1^2}{w^2}\right) \right] \, .
\end{split}
\label{eq:hybr_gauss_in_f}
\end{equation}
Here $0 < \epsilon \ll 1$ is a small dimensionless parameter, and $x_1 = \vec{r}_\perp \cdot \hat{\vec{e}}_1$. The width $w$ of the wave packet is taken
much larger than the two healing lengths $(\beta_0 g_{d,s} I_0)^{-1/2}$. We will see below that this condition makes the present setup well-suited to study
the low-$q_\perp$ part of the Bogoliubov spectrum, i.e., the sound modes. The components of the polarization vector of the fluctuation are controlled by
the angle $\vartheta_f$. In terms of the density-phase variables, the incident state~\eqref{eq:hybr_gauss_in_f} corresponds to
\begin{subequations}
\label{eq:hybr_gauss_in_dp}
\begin{align}
\delta I(\vec{r}_\perp,z=0) &{} = 2 \epsilon I_0 \cos\frac{\Delta\vartheta}{2} \exp\left(- \frac{x_1^2}{w^2}\right) \, ,
\label{eq:hybr_gauss_in_int} \\
\delta \vartheta(\vec{r}_\perp,z=0) &{} = 2 \epsilon \sin\frac{\Delta\vartheta}{2} \exp\left(- \frac{x_1^2}{w^2}\right) \, ,
\label{eq:hybr_gauss_in_pol} \\
\Theta(\vec{r}_\perp,z=0) &{} = \delta\chi(\vec{r}_\perp,z=0) = 0 \, ,
\label{eq:hybr_gauss_in_ph}
\end{align}
\end{subequations}
where $\Delta\vartheta = \vartheta_f - \pi/2$. Taking the Fourier transform of Eqs.~\eqref{eq:hybr_gauss_in_dp}, one obtains the $z=0$ value
of the four-component vector $X$,
\begin{equation}
\begin{split}
&{} X(\vec{q}_\perp,z=0) \\
&{} =
\epsilon
\begin{pmatrix}
\cos\frac{\Delta\vartheta}{2} \\
\sin\frac{\Delta\vartheta}{2} \\
0 \\
0
\end{pmatrix}
2 \pi \sqrt{\pi w^2} \exp\left( - \frac{w^2 q_1^2}{4} \right) \delta(q_2) \, ,
\end{split}
\label{eq:hybr_gauss_in_X}
\end{equation}
with $q_{1(2)} = \vec{q}_\perp \cdot \hat{\vec{e}}_{1(2)}$. We subsequently take $\Pi^*(\vec{q}_\perp,z=0)$ as in Eq.~\eqref{eq:cm_init_cond_Pi},
so that only the low-frequency Bogoliubov modes are excited. The weights of such modes are given by Eq.~\eqref{eq:dp_weight} and can be written
in the form
\begin{equation}
C_\ell(\vec{q}_\perp) = \epsilon \tilde{C}_\ell(\vec{q}_\perp)
2\pi \sqrt{\pi w^2} \exp\left(- \frac{w^2 q_1^2}{4}\right) \delta(q_2) \, .
\label{eq:hybr_gauss_weights}
\end{equation}
The exact expressions of the $\tilde{C}_\ell(\vec{q}_\perp)$'s depend on those of the amplitudes~\eqref{eq:dp_norm_amp_X}. However, in the next
steps of the calculation only their behavior at low $q_\perp$ will be needed. We now insert the weights~\eqref{eq:hybr_gauss_weights} and
amplitudes~\eqref{eq:dp_norm_amp_X} into the superposition~\eqref{eq:dp_ham_sol} and take the inverse Fourier transform. In carrying out the
calculations we exploit the fact that the Gaussian in Eq.~\eqref{eq:hybr_gauss_weights} is very
narrow, thus one can approximate $\Omega_{-,d(s)}(q_1,q_2=0) \simeq c_{d(s)}(\varphi_f) q_1$. Besides one can prove that $\tilde{C}_{-,d}
\delta\tilde{I}_{-,d} \to \cos(\Delta\vartheta/2)/2$ and $\tilde{C}_{-,s} \delta\tilde{I}_{-,s} \to \sin(\Delta\vartheta/2)/2$ as $q_\perp \to 0$, while
both $\tilde{C}_{-,s} \delta\tilde{I}_{-,d}$ and $\tilde{C}_{-,d} \delta\tilde{I}_{-,s}$ vanish in this limit. Using these results one eventually obtains
\begin{align}
\begin{split}
\delta I(\vec{r}_\perp,z) = {}&{} \epsilon I_0 \cos\frac{\Delta\vartheta}{2}
\Bigg\{ \exp\left[- \frac{\left( x_1 - c_d z \right)^2}{w^2} \right] \\
&{} + \exp\left[- \frac{\left( x_1 + c_d z \right)^2}{w^2} \right] \Bigg\} \, ,
\end{split}
\label{eq:hybr_gauss_int} \\
\begin{split}
\delta \vartheta(\vec{r}_\perp,z) = {}&{} \epsilon \sin\frac{\Delta\vartheta}{2}
\Bigg\{ \exp\left[- \frac{\left( x_1 - c_s(\varphi_f) z \right)^2}{w^2} \right] \\
&{} + \exp\left[- \frac{\left( x_1 + c_s(\varphi_f) z \right)^2}{w^2} \right] \Bigg\} \, ,
\end{split}
\label{eq:hybr_gauss_pol}
\end{align}
where $c_d$ and $c_s$ are given by Eqs.~\eqref{eq:dp_sound_vel_d} and~\eqref{eq:dp_sound_vel_s}, respectively. These equations describe the emission
of pairs of Bogoliubov quasiparticles from $z=0$ onward. They show that if the fluctuation at the interface has the same polarization as the background, i.e.,
$\Delta\vartheta = 0$, only the density sound mode is excited. The latter was observed in the experiment of Ref.~\cite{Fontaine2018} through measurements
of the total intensity, $I=I_0+\delta I$. In order to excite the spin sound mode only, one has to choose $\Delta\vartheta = \pm \pi$. This corresponds to a
fluctuation polarized along the $\pm \mi \hat{\vec{e}}_y$ directions, i.e., perpendicular to the background and with a phase difference of $\pm \pi/2$. 
The spin sound mode can be detected by measuring the relative intensity of the two polarization components, $|\mathcal{E}_+|^2 - |\mathcal{E}_-|^2 =
- I_0 \delta \vartheta$. If one measures $|\mathcal{E}_+|^2$ and $|\mathcal{E}_-|^2$ separately, one finds that they split into four Gaussian branches, having
different weights and propagating with velocities $\pm c_d$ and $\pm c_s$. 

We stress that Eqs.~\eqref{eq:hybr_gauss_int} and~\eqref{eq:hybr_gauss_pol} do not assume any paraxial approximation. In the paraxial regime, they keep
the same form, with the sound velocities replaced by their paraxial counterparts. In particular, in both the paraxial and nonparaxial descriptions the two sound
modes keep a pure density and spin nature, irrespective of the propagation direction fixed by $\varphi_f$. In the next section we will see that this is no longer
true when larger-$q_\perp$ modes are excited.

\subsection{Phase shift experiment}
\label{subsec:hybr_phase_exp}
A drawback of the previous configuration is the difficulty to excite modes whose wavelength is of the order or larger than the healing lengths
$(\beta_0 g_{d,s} I_0)^{-1/2}$. Such higher-$q_\perp$ modes, however, are precisely those expected to display signatures of spin-orbit coupling. To circumvent
this issue, a possibility would be to consider a wave packet carrying a finite mean momentum $\vec{k}_1 $ such that $k_1 w \gg 1$. For simplicity however, here
we restrict ourselves to the case of a pure phase perturbation of momentum $\vec{k}_1$, which enables one to excite individual modes of given wavenumber
$\vec{q}_\perp = \vec{k}_1$. The form of such an input field is
\begin{equation}
\begin{split}
&{} \begin{pmatrix}
\mathcal{E}_+(\vec{r}_\perp,z=0) \\
\mathcal{E}_-(\vec{r}_\perp,z=0)
\end{pmatrix} \\
&{} = \sqrt{I_0} \left[
\begin{pmatrix}
- 1 / \sqrt{2} \\
1 / \sqrt{2}
\end{pmatrix}
+ \epsilon
\begin{pmatrix}
- \cos\frac{\vartheta_f}{2} \\
\sin\frac{\vartheta_f}{2}
\end{pmatrix}
\exp\left(\mi k_1 x_1\right) \right] \, .
\end{split}
\label{eq:hybr_phase_in_f}
\end{equation}
This is the same as Eq.~\eqref{eq:hybr_gauss_in_f}, except that the Gaussian is replaced by a plane wave with wave vector $\vec{k}_1 = k_1 \hat{\vec{e}}_1$,
see Fig.~\ref{fig:hybr_configs}(b). The corresponding fluctuations of the density-phase variables are
\begin{subequations}
\label{eq:hybr_phase_in_dp}
\begin{align}
\delta I(\vec{r}_\perp,z=0) &{} = 2 \epsilon I_0 \cos\frac{\Delta\vartheta}{2} \cos(k_1 x_1) \, ,
\label{eq:hybr_phase_in_int} \\
\delta \vartheta(\vec{r}_\perp,z=0) &{} = 2 \epsilon \sin\frac{\Delta\vartheta}{2} \cos(k_1 x_1) \, ,
\label{eq:hybr_phase_in_pol} \\
\Theta(\vec{r}_\perp,z=0) &{} = \epsilon \cos\frac{\Delta\vartheta}{2} \sin(k_1 x_1) \, ,
\label{eq:hybr_phase_in_gp} \\
\delta\chi(\vec{r}_\perp,z=0) &{} = - 2 \epsilon \sin\frac{\Delta\vartheta}{2} \sin (k_1 x_1) \, .
\label{eq:hybr_phase_in_rp}
\end{align}
\end{subequations}
The input value of the vector $X$ can then be written as $X(\vec{q}_\perp,z=0) = X_\delta(\vec{q}_\perp) + X_\delta^*(-\vec{q}_\perp)$, where
\begin{equation}
X_\delta(\vec{q}_\perp) =
2 \epsilon \, \pi^2
\begin{pmatrix}
\cos\frac{\Delta\vartheta}{2} \\
\sin\frac{\Delta\vartheta}{2} \\
- \mi \cos\frac{\Delta\vartheta}{2} \\
\mi \sin\frac{\Delta\vartheta}{2}
\end{pmatrix}
\delta(\vec{q}_\perp - \vec{k}_1) \, ,
\label{eq:hybr_phase_in_X}
\end{equation}
and that of $\Pi^*$ is given by Eq.~\eqref{eq:cm_init_cond_Pi}. The weights of the various modes can be computed using Eq.~\eqref{eq:dp_weight},
or Eq.~\eqref{eq:paraxial_dp_weight_r} in the paraxial framework. They are given by a linear combination of $\delta(\vec{q}_\perp-\vec{k}_1)$ and
$\delta(\vec{q}_\perp+\vec{k}_1)$. Again we insert these weights and the amplitudes~\eqref{eq:dp_norm_amp_X} into the general
expression~\eqref{eq:dp_ham_sol} and invert the Fourier transform. In the paraxial approximation, this procedure yields the results
\begin{align}
\begin{split}
\delta I(\vec{r}_\perp,z) = {}&{} 2 \epsilon I_0 \cos\frac{\Delta\vartheta}{2}
\Big\{ \lambda_d^{(+)}(\vec{k}_1) \cos[k_1 x_1 - \Omega_d(\vec{k}_1) z] \\
&{} + \lambda_d^{(-)}(\vec{k}_1) \cos[k_1 x_1 + \Omega_d(\vec{k}_1) z] \Big\} \, ,
\end{split}
\label{eq:hybr_phase_out_int_par} \\
\begin{split}
\delta \vartheta(\vec{r}_\perp,z) = {}&{} 2 \epsilon \sin\frac{\Delta\vartheta}{2}
\Big\{ \lambda_s^{(+)}(\vec{k}_1) \cos[k_1 x_1 - \Omega_s(\vec{k}_1) z] \\
&{} + \lambda_s^{(-)}(\vec{k}_1) \cos[k_1 x_1 + \Omega_s(\vec{k}_1) z] \Big\} \, ,
\end{split}
\label{eq:hybr_phase_out_pol_par}
\end{align}
where $\lambda_\ell^{(\pm)}(\vec{k}_1) = [\Omega_\ell(\vec{k}_1) \pm k_1^2/2\beta_0] / 2 \Omega_\ell(\vec{k}_1)$.
Hence, the total and relative intensity oscillate in effective time at a single frequency, equal to $\Omega_d(\vec{k}_1)$ and $\Omega_s(\vec{k}_1)$, respectively.
As in the small Gaussian probe experiment, when the background and fluctuation input fields have equal polarization ($\Delta\vartheta = 0$) the sole density
mode is excited. The experiment of Ref.~\cite{Fontaine2018} was performed in such conditions. In more general situations one has to take the existence of
the spin mode into account. This effect becomes more and more relevant as $\Delta\vartheta$ increases. In particular, when $\Delta\vartheta = \pm \pi$
only the spin oscillation is visible, again in agreement with the findings of Sec.~\ref{subsec:hybr_gauss_exp}.

\begin{figure}
\includegraphics[scale=1]{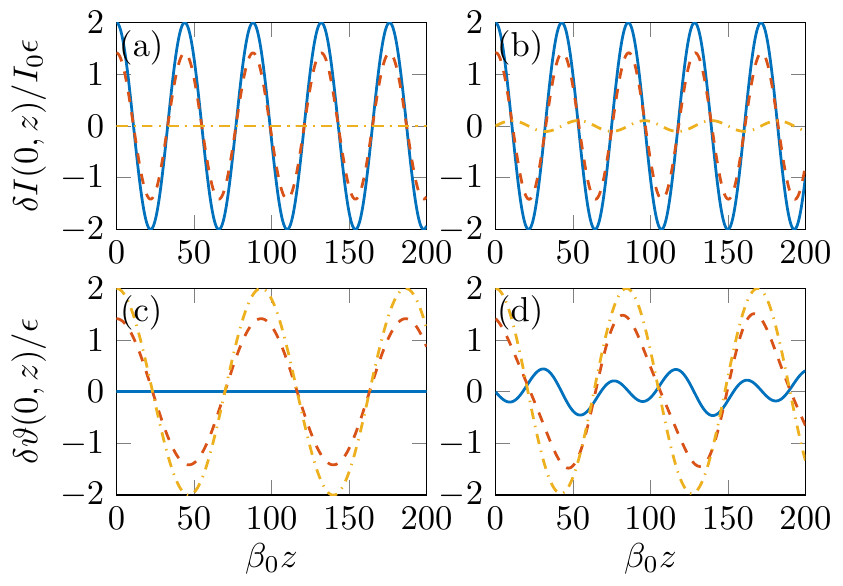}
\caption{Fluctuations of (a)-(b) the total and (c)-(d) the relative optical intensity at $x_1 = 0$ as functions of the effective time $z$. 
We take $\varphi_f=0$ in (a) and (c) (no mode hybridization), and $\varphi_f=\pi/4$ in (b) and (d) (mode hybridization is present). In each panel we show
the results for $\Delta\vartheta = 0$ (blue solid line), $\pi/2$ (red dashed line), and $\pi$ (yellow dash-dot line). Observe that in panel (b) [(d)], the fluctuation of
the total (relative) intensity oscillates when $\Delta\vartheta = \pi$ ($\Delta\vartheta = 0$), whereas it strictly vanishes in case (a) [(c)]. This phenomenon is a marked
signature of spin-orbit coupling. For these plots we choose the magnitude of the fluctuation wave vector equal to $k_1 = 0.2\,\beta_0$ and the same nonlinear
couplings $g_d I_0 / \beta_0 = 0.2$ and $g_s I_0 / \beta_0 = 0.05$ as Fig.~\ref{fig:dp_bogo_sp}.}
\label{fig:hybr_ph_sh}
\end{figure}

A novel phenomenon arises when one describes the problem on the basis of the general formalism presented in Secs.~\ref{sec:model}, \ref{sec:dens_phase},
and~\ref{sec:bogo_theory}, i.e., without resorting to the paraxial approximation. In this case, we find that the total and relative intensities have the structure
\begin{widetext}
\begin{align}
\begin{split}
\delta I(\vec{r}_\perp,z) =
{}&{} 2 \epsilon I_0 \cos\frac{\Delta\vartheta}{2} \sum_{\ell = d, s}
\Big\{ \lambda_{I,\ell}^{(+)}(\vec{k}_1) \cos[k_1 x_1 - \Omega_{-,\ell}(\vec{k}_1) z]
+ \lambda_{I,\ell}^{(-)}(-\vec{k}_1) \cos[k_1 x_1 + \Omega_{-,\ell}(-\vec{k}_1) z] \Big\} \\
{}&{} - 2 \epsilon I_0 \sin\frac{\Delta\vartheta}{2} \sum_{\ell = d, s}
\Big\{ \mu_{I,\ell}^{(+)}(\vec{k}_1) \sin[k_1 x_1 - \Omega_{-,\ell}(\vec{k}_1) z]
+ \mu_{I,\ell}^{(-)}(-\vec{k}_1) \sin[k_1 x_1 + \Omega_{-,\ell}(-\vec{k}_1) z] \Big\} \, ,
\end{split}
\label{eq:hybr_phase_out_int} \\
\begin{split}
\delta \vartheta(\vec{r}_\perp,z) =
{}&{} 2 \epsilon \sin\frac{\Delta\vartheta}{2} \sum_{\ell = d, s}
\Big\{ \lambda_{\vartheta,\ell}^{(+)}(\vec{k}_1) \cos[k_1 x_1 - \Omega_{-,\ell}(\vec{k}_1) z]
+ \lambda_{\vartheta,\ell}^{(-)}(-\vec{k}_1) \cos[k_1 x_1 + \Omega_{-,\ell}(-\vec{k}_1) z] \Big\} \\
{}&{} - 2 \epsilon \cos\frac{\Delta\vartheta}{2} \sum_{\ell = d, s}
\Big\{ \mu_{\vartheta,\ell}^{(+)}(\vec{k}_1) \sin[k_1 x_1 - \Omega_{-,\ell}(\vec{k}_1) z]
+ \mu_{\vartheta,\ell}^{(-)}(-\vec{k}_1) \sin[k_1 x_1 + \Omega_{-,\ell}(-\vec{k}_1) z] \Big\} \, .
\end{split}
\label{eq:hybr_phase_out_pol} 
\end{align}
\end{widetext}
The values of the coefficients $\smash{\lambda_{I,\ell}^{(\pm)}}$, $\smash{\lambda_{\vartheta,\ell}^{(\pm)}}$, $\smash{\mu_{I,\ell}^{(\pm)}}$,
and $\smash{\mu_{\vartheta,\ell}^{(\pm)}}$, which depend on the amplitudes~\eqref{eq:dp_norm_amp_X}, have to be computed numerically in general.
Only in the paraxial regime of small momenta $(k_1/\beta_0)^2 \ll 1$ and weak interactions $|g_{d,s}| I_0 / \beta_0 \ll 1$ they reduce to a simple analytical
expression, given by the coefficients in Eqs.~\eqref{eq:hybr_phase_out_int_par}--\eqref{eq:hybr_phase_out_pol_par}.

Here one has to distinguish two cases. If $\vec{k}_1$ is parallel or perpendicular to the $x$ axis, i.e., $\varphi_f=0,\pi/2,\pi$, then $\smash{\lambda_{I,s}^{(\pm)}
= \mu_{I,s}^{(\pm)} = 0}$ and $\smash{\lambda_{\vartheta,d}^{(\pm)} = \mu_{\vartheta,d}^{(\pm)} = 0}$. Hence, the scenario is qualitatively similar to the paraxial
one, where $\delta I$ and $\delta\vartheta$ oscillate at a single frequency. This is illustrated in Figs.~\ref{fig:hybr_ph_sh}(a) and~\ref{fig:hybr_ph_sh}(c),
where we plot the time evolution of these two quantities evaluated at $x_1 = 0$. 

A dramatic change occurs when $\vec{k}_1$ is not oriented along one of the special directions mentioned above. In this case, one observes a phenomenon
of mode hybridization, as $\delta I$ and $\delta\vartheta$ oscillate at both frequencies $\Omega_{-,d}(\vec{k}_1)$ and $\Omega_{-,s}(\vec{k}_1)$.
This is depicted in Figs.~\ref{fig:hybr_ph_sh}(b) and~\ref{fig:hybr_ph_sh}(d), where the oscillations are characterized by a beat that is particularly
visible in the behavior of $\delta\vartheta(0,z)$. Another remarkable feature is that both $\delta I$ and $\delta\vartheta$ oscillate even when
$\Delta\vartheta = 0,\pm \pi$, revealing once more the hybrid nature of the $d$ and $s$ modes. Mode hybridization is one of the most striking
consequences of the spin-orbit coupling. A similar phenomenon of beat between two hybrid modes was recently studied in the context of trapped
atomic spin-orbit-coupled Bose-Einstein condensates, where it can be observed when the gas is in the supersolid phase~\cite{Geier2021}. Our
setting provides an alternative and viable way to study the impact of spin-orbit coupling on the collective modes of an interacting optical system.

\section{Conclusion}
\label{sec:conclusion}
We have developed a density-phase, Bogoliubov formalism for studying the propagation of light in a bulk Kerr nonlinear medium. Unlike usual approaches
employed to address this problem, our formalism is general and does not rely on the paraxial approximation. Within this framework, we have derived the
Bogoliubov equations governing the effective-time evolution of fluctuations on top of a linearly polarized background field. By solving these equations, we
have obtained the frequencies and amplitudes of the Bogoliubov modes, whose formal properties have been carefully investigated. The Bogoliubov spectrum 
is made of several anisotropic branches, each associated with a density- or spin-like oscillation of the two polarization components of the light. For slowly
varying electric fields and weak nonlinearities, one recovers the results of the paraxial approximation, for which the fluid of light behaves like an out-of-equilibrium
binary Bose mixture. 

Our description has also allowed us to reveal the existence of a mechanism of spin-orbit coupling arising beyond the paraxial approximation. This phenomenon,
naturally present in inhomogeneous media, here manifests itself in the context of a weakly inhomogeneous fluid of light, where a small fluctuation of the fluid
couples to the light polarization to modify the fluid properties. In particular, it leads to an anisotropy of the Bogoliubov spectrum and to the hybridization of the
density and spin modes. By finally investigating a simple experimental protocol involving a probed beam sent through a nonlinear medium, we have proposed
a simple strategy to 1) separately investigate the density and the so far never measured spin mode, and 2) detect the hybridization of these modes by optical
spin-orbit coupling.

Our results pave the way to future studies on nonparaxial effects, spin-orbit coupling and related chiral phenomena~\cite{Petersen2014,Lodahl2017} in cavityless
fluids of light. In this context, a natural extension of our work would be to study the interplay between nonparaxial effects and the phenomenon of nonlinear
birefringence, which occurs when the background field is elliptically polarized~\cite{Agrawal_book}. It would be equally interesting to consider fields propagating
at an angle with respect to the optical axis, which would enable one to characterize light superfluidity and its interplay with spin-orbit coupling. In atomic gases
such an interplay gives rise to novel configurations, including quantum phases with supersolid features~\cite{Li2013,Li2017}. Implemented in a context of optical
fluid mixtures, it might open a path to the possible phenomenon of light supersolidity.

\begin{acknowledgments}
We thank A. Bramati, I. Carusotto, and N. Pavloff for fruitful discussions. 
We acknowledge financial support from the Agence Nationale de la Recherche (grant ANR-19-CE30-0028-01 CONFOCAL).
\end{acknowledgments}

\appendix

\section{Coefficients of the kinetic Lagrangian}
\label{app:Kcoeff}
The coefficients of the transverse Lagrangian~\eqref{eq:dp_lagr_perp} are given by:
\begin{widetext}
\begin{align*}
(K_{II})_{\alpha\alpha'} = {}&{} (K_{\Theta\Theta})_{\alpha\alpha'} =
 (S_\alpha S_{\alpha'})_{++} \cos^2\frac{\vartheta}{2} + (S_\alpha S_{\alpha'})_{--} \sin^2\frac{\vartheta}{2}
+ (S_\alpha S_{\alpha'})_{+-} \frac{\sin\vartheta}{2} \, \me^{- \mi \chi} + (S_\alpha S_{\alpha'})_{-+} \frac{\sin\vartheta}{2} \, \me^{\mi \chi} \, , \\
(K_{\vartheta\vartheta})_{\alpha\alpha'} = {}&{} (K_{\chi\chi})_{\alpha\alpha'} =
 (S_\alpha S_{\alpha'})_{++} \sin^2\frac{\vartheta}{2} + (S_\alpha S_{\alpha'})_{--} \cos^2\frac{\vartheta}{2}
- (S_\alpha S_{\alpha'})_{+-} \frac{\sin\vartheta}{2} \, \me^{- \mi \chi} - (S_\alpha S_{\alpha'})_{-+} \frac{\sin\vartheta}{2} \, \me^{\mi \chi} \, , \\
(K_{I\vartheta})_{\alpha\alpha'} = {}&{}
- 2 \Re \bigg[ (S_\alpha S_{\alpha'})_{++} \frac{\sin\vartheta}{2} - (S_\alpha S_{\alpha'})_{--} \frac{\sin\vartheta}{2}
- (S_\alpha S_{\alpha'})_{+-} \cos^2\frac{\vartheta}{2} \, \me^{- \mi \chi} + (S_\alpha S_{\alpha'})_{-+} \sin^2\frac{\vartheta}{2} \, \me^{\mi \chi} \bigg] \, , \\
(K_{I\Theta})_{\alpha\alpha'} = {}&{}
- 2 \Im \bigg[ (S_\alpha S_{\alpha'})_{++} \cos^2\frac{\vartheta}{2} + (S_\alpha S_{\alpha'})_{--} \sin^2\frac{\vartheta}{2}
+ (S_\alpha S_{\alpha'})_{+-} \frac{\sin\vartheta}{2} \, \me^{- \mi \chi} + (S_\alpha S_{\alpha'})_{-+} \frac{\sin\vartheta}{2} \, \me^{\mi \chi} \bigg] \, , \\
(K_{I\chi})_{\alpha\alpha'} = {}&{}
- 2 \Im \bigg[ (S_\alpha S_{\alpha'})_{++} \cos^2\frac{\vartheta}{2} - (S_\alpha S_{\alpha'})_{--} \sin^2\frac{\vartheta}{2}
- (S_\alpha S_{\alpha'})_{+-} \frac{\sin\vartheta}{2} \, \me^{- \mi \chi} + (S_\alpha S_{\alpha'})_{-+} \frac{\sin\vartheta}{2} \, \me^{\mi \chi} \bigg] \, , \\
(K_{\vartheta\Theta})_{\alpha\alpha'} = {}&{}
2 \Im \bigg[ (S_\alpha S_{\alpha'})_{++} \frac{\sin\vartheta}{2} - (S_\alpha S_{\alpha'})_{--} \frac{\sin\vartheta}{2}
+ (S_\alpha S_{\alpha'})_{+-} \sin^2\frac{\vartheta}{2} \, \me^{- \mi \chi} - (S_\alpha S_{\alpha'})_{-+} \cos^2\frac{\vartheta}{2} \, \me^{\mi \chi} \bigg] \, , \\
(K_{\vartheta\chi})_{\alpha\alpha'} = {}&{}
2 \Im \bigg[ (S_\alpha S_{\alpha'})_{++} \frac{\sin\vartheta}{2} + (S_\alpha S_{\alpha'})_{--} \frac{\sin\vartheta}{2}
- (S_\alpha S_{\alpha'})_{+-} \sin^2\frac{\vartheta}{2} \, \me^{- \mi \chi} - (S_\alpha S_{\alpha'})_{-+} \cos^2\frac{\vartheta}{2} \, \me^{\mi \chi} \bigg] \, , \\
(K_{\Theta\chi})_{\alpha\alpha'} = {}&{}
2 \Re \bigg[ (S_\alpha S_{\alpha'})_{++} \cos^2\frac{\vartheta}{2} - (S_\alpha S_{\alpha'})_{--} \sin^2\frac{\vartheta}{2}
- (S_\alpha S_{\alpha'})_{+-} \frac{\sin\vartheta}{2} \, \me^{- \mi \chi} + (S_\alpha S_{\alpha'})_{-+} \frac{\sin\vartheta}{2} \, \me^{\mi \chi} \bigg] \, .
\end{align*}
\end{widetext}

\section{Expression of the matrices in the Bogoliubov Lagrangian}
\label{app:bogo_mat}
In this appendix we provide the expressions of the three matrices entering the Bogoliubov Lagrangian density~\eqref{eq:dp_lagr_f}. Such matrices
have the following structure $(k=0,1,2)$:
\begin{equation*}
\Lambda_k = I_0
\begin{pmatrix}
(\Lambda_k)_{1,1} & (\Lambda_k)_{1,2} & (\Lambda_k)_{1,3} & (\Lambda_k)_{1,4} \\
(\Lambda_k)_{2,1} & (\Lambda_k)_{2,2} & (\Lambda_k)_{2,3} & (\Lambda_k)_{2,4} \\
(\Lambda_k)_{3,1} & (\Lambda_k)_{3,2} & (\Lambda_k)_{3,3} & (\Lambda_k)_{3,4} \\
(\Lambda_k)_{4,1} & (\Lambda_k)_{4,2} & (\Lambda_k)_{4,3} & (\Lambda_k)_{4,4} \\
\end{pmatrix} \, .
\end{equation*}
Hence, each matrix has 16 entries, but we find that half of them are zero. The nonvanishing entries of $\Lambda_0$ are
\begin{align*}
&{} (\Lambda_0)_{1,1} =
\begin{aligned}[t]
&{} \frac{q_\perp^2}{4\beta_0} \left[ (1 - \cos 2\varphi)
- \frac{k^2 \left(1 + \cos 2\varphi \right)}{q_\perp^2 - (k_0^2 - \Delta k_0^2)} \right] \\
&{} + 2 g_d I_0 \, ,
\end{aligned} \\
&{} (\Lambda_0)_{1,4} = (\Lambda_0)_{4,1} =
\frac{q_\perp^2}{4\beta_0} \sin 2\varphi
\left[ 1 + \frac{k^2}{q_\perp^2 - (k_0^2 - \Delta k_0^2)} \right] \, , \\
&{} (\Lambda_0)_{2,2} =
\begin{aligned}[t]
&{} \frac{q_\perp^2}{4\beta_0} \left[ (1 + \cos 2\varphi)
- \frac{k^2 \left(1 - \cos 2\varphi \right)}{q_\perp^2 - (k_0^2 + \Delta k_0^2)} \right] \\
&{} + 2 g_s I_0 \, ,
\end{aligned} \\
&{} (\Lambda_0)_{2,3} = (\Lambda_0)_{3,2} =
\frac{q_\perp^2}{4\beta_0} \sin 2\varphi
\left[ 1 + \frac{k^2}{q_\perp^2 - (k_0^2 + \Delta k_0^2)} \right] \, , \\
&{} (\Lambda_0)_{3,3} =
\frac{q_\perp^2}{4\beta_0} \left[ (1 - \cos 2\varphi)
- \frac{k^2 \left(1 + \cos 2\varphi \right)}{q_\perp^2 - (k_0^2 + \Delta k_0^2)} \right] \, , \\
&{} (\Lambda_0)_{4,4} =
\frac{q_\perp^2}{4\beta_0} \left[ (1 + \cos 2\varphi)
- \frac{k^2 \left(1 - \cos 2\varphi \right)}{q_\perp^2 - (k_0^2 - \Delta k_0^2)} \right] \, .
\end{align*}
The nonzero entries of $\Lambda_1$ are
\begin{align*}
&{} (\Lambda_1)_{1,2} = (\Lambda_1)_{4,3} = \frac{q_\perp^2}{4\beta_0} \frac{k \sin 2\varphi}{q_\perp^2 - (k_0^2 + \Delta k_0^2)} \, , \\
&{} (\Lambda_1)_{1,3} = - \frac{q_\perp^2}{4\beta_0} \frac{k \left(1 + \cos 2\varphi\right)}{q_\perp^2 - (k_0^2 + \Delta k_0^2)} \, , \\
&{} (\Lambda_1)_{2,1} = (\Lambda_1)_{3,4} = - \frac{q_\perp^2}{4\beta_0} \frac{k \sin 2\varphi}{q_\perp^2 - (k_0^2 - \Delta k_0^2)} \, , \\
&{} (\Lambda_1)_{2,4} = \frac{q_\perp^2}{4\beta_0} \frac{k \left(1 - \cos 2\varphi\right)}{q_\perp^2 - (k_0^2 - \Delta k_0^2)} \, , \\
&{} (\Lambda_1)_{3,1} = - \frac{k}{\beta_0} + \frac{q_\perp^2}{4\beta_0} \frac{k \left(1 + \cos 2\varphi\right)}{q_\perp^2 - (k_0^2 - \Delta k_0^2)} \, , \\
&{} (\Lambda_1)_{4,2} = \frac{k}{\beta_0} - \frac{q_\perp^2}{4\beta_0} \frac{k \left(1 - \cos 2\varphi\right)}{q_\perp^2 - (k_0^2 + \Delta k_0^2)} \, .
\end{align*}
Finally, the nonvanishing entries of $\Lambda_2$ are
\begin{align*}
&{} (\Lambda_2)_{1,1} = - \frac{1}{2\beta_0} + \frac{q_\perp^2}{4\beta_0} \frac{1 + \cos 2\varphi}{q_\perp^2 - (k_0^2 + \Delta k_0^2)} \, , \\
&{} (\Lambda_2)_{1,4} = (\Lambda_2)_{4,1} = - \frac{q_\perp^2}{4\beta_0} \frac{\sin 2\varphi}{q_\perp^2 - (k_0^2 + \Delta k_0^2)^2} \, , \\
&{} (\Lambda_2)_{2,2} = - \frac{1}{2\beta_0} + \frac{q_\perp^2}{4\beta_0} \frac{1 - \cos 2\varphi}{q_\perp^2 - (k_0^2 - \Delta k_0^2)} \, , \\
&{} (\Lambda_2)_{2,3} = (\Lambda_2)_{3,2} = - \frac{q_\perp^2}{4\beta_0} \frac{\sin 2\varphi}{q_\perp^2 - (k_0^2 - \Delta k_0^2)} \, , \\
&{} (\Lambda_2)_{3,3} = - \frac{1}{2\beta_0} + \frac{q_\perp^2}{4\beta_0} \frac{1 + \cos 2\varphi}{q_\perp^2 - (k_0^2 - \Delta k_0^2)} \, , \\
&{} (\Lambda_2)_{4,4} = - \frac{1}{2\beta_0} + \frac{q_\perp^2}{4\beta_0} \frac{1 - \cos 2\varphi}{q_\perp^2 - (k_0^2 + \Delta k_0^2)} \, . 
\end{align*}

\section{Conjugate momenta at the interface}
\label{app:conj_mom_int}
In this appendix we explain how one can choose the interface value $\Pi^*(\vec{q}_\perp,z\!=\!0)$ of the conjugate momenta such that
the upper Bogoliubov modes, having frequency $\Omega_{+,d}$ and $\Omega_{+,s}$, remain unpopulated. The procedure is the
following. First, we consider the matrix $\mathcal{P}$ bringing $\mathcal{B}$ into diagonal form,
\begin{equation}
\mathcal{P}^{-1} \mathcal{B} \mathcal{P} = \mathcal{B}_D =
\begin{pmatrix}
\mathcal{B}_{D+} & 0 \\
0 & \mathcal{B}_{D-}
\end{pmatrix} \, ,
\label{eq:cm_diag_mat}
\end{equation}
where
\begin{equation}
\begin{split}
\mathcal{B}_{D\pm} = &{} \operatorname{diag}
\big( \Omega_{d,\pm}(\vec{q}_\perp), -\Omega_{d,\pm}^*(-\vec{q}_\perp), \Omega_{s,\pm}(\vec{q}_\perp), \\
&{} -\Omega_{s,\pm}^*(-\vec{q}_\perp) \big) \, .
\end{split}
\label{eq:cm_mat_block_B}
\end{equation}
$\mathcal{P}$ and $\mathcal{P}^{-1}$ are $8 \times 8$ matrices that can be split into several $4 \times 4$ blocks,
\begin{equation}
\mathcal{P} =
\begin{pmatrix}
\mathcal{P}_{X+} & \mathcal{P}_{X-} \\
\mathcal{P}_{\Pi+} & \mathcal{P}_{\Pi-}
\end{pmatrix} \, , \;\;\;
\mathcal{P}^{-1} =
\begin{pmatrix}
\left(\mathcal{P}^{-1}\right)_{X+} & \left(\mathcal{P}^{-1}\right)_{\Pi+} \\
\left(\mathcal{P}^{-1}\right)_{X-} & \left(\mathcal{P}^{-1}\right)_{\Pi-}
\end{pmatrix} \, .
\label{eq:cm_mat_block_P}
\end{equation}
The columns of $\mathcal{P}_{X\pm}$ ($\mathcal{P}_{\Pi\pm}$) coincide with the amplitudes $X_{0,\ell}$ ($\Pi_{0,\ell}^*$) written
in the proper order. We now define the new variables
\begin{equation}
\begin{pmatrix}
Y_{D+} \\
Y_{D-}
\end{pmatrix}
= \mathcal{P}^{-1}
\begin{pmatrix}
X \\
\Pi
\end{pmatrix} \, ,
\label{eq:cm_def_Y}
\end{equation}
whose effective-time evolution is trivial, 
\begin{equation}
Y_{D\pm}(\vec{q}_\perp,z) = \me^{- \mi \mathcal{B}_{D\pm} z} Y_{D\pm}(\vec{q}_\perp,z\!=\!0) \, .
\label{eq:cm_evol_Y}
\end{equation}
Inverting Eq.~\eqref{eq:cm_def_Y} we can express the original variables $X$ and $\Pi$ as combinations of $Y_{D+}$ and $Y_{D-}$. In order not to excite
the high-frequency modes one has to impose the condition $Y_{D+}(\vec{q}_\perp,z\!=\!0) = 0$, that is,
\begin{equation}
\Pi^*(\vec{q}_\perp,z\!=\!0) = - \left[ \left(\mathcal{P}^{-1}\right)_{\Pi+} \right]^{-1} \left(\mathcal{P}^{-1}\right)_{X+} X(\vec{q}_\perp,z\!=\!0) \, .
\label{eq:cm_init_cond_Pi}
\end{equation}

\end{document}